\documentclass{egpubl}
\usepackage{MAM2023}

 \electronicVersion 

\ifpdf \usepackage[pdftex]{graphicx} \pdfcompresslevel=9
\else \usepackage[dvips]{graphicx} \fi

\PrintedOrElectronic

\usepackage{t1enc,dfadobe}

\usepackage{egweblnk}
\usepackage{cite}

\usepackage{graphicx}
\usepackage[permil]{overpic}
\usepackage[export]{adjustbox}
\usepackage{pict2e} 
\usepackage{multirow}
\usepackage{xcolor}
\usepackage{booktabs} 
\usepackage{amssymb}
\usepackage{amsfonts}
\usepackage{amsmath}
\usepackage[scaled=.90]{helvet}
\usepackage[normalem]{ulem}
\usepackage{subcaption}

\title[Neural Texture Block Compression]%
      {Neural Texture Block Compression}

\author[S. Fujieda \& T. Harada]
       {S. Fujieda
        and T. Harada
        \\
         Advanced Micro Devices, Inc.
       }

\begin{document}

\teaser{
  \centering
  \includegraphics[width=0.9\linewidth]{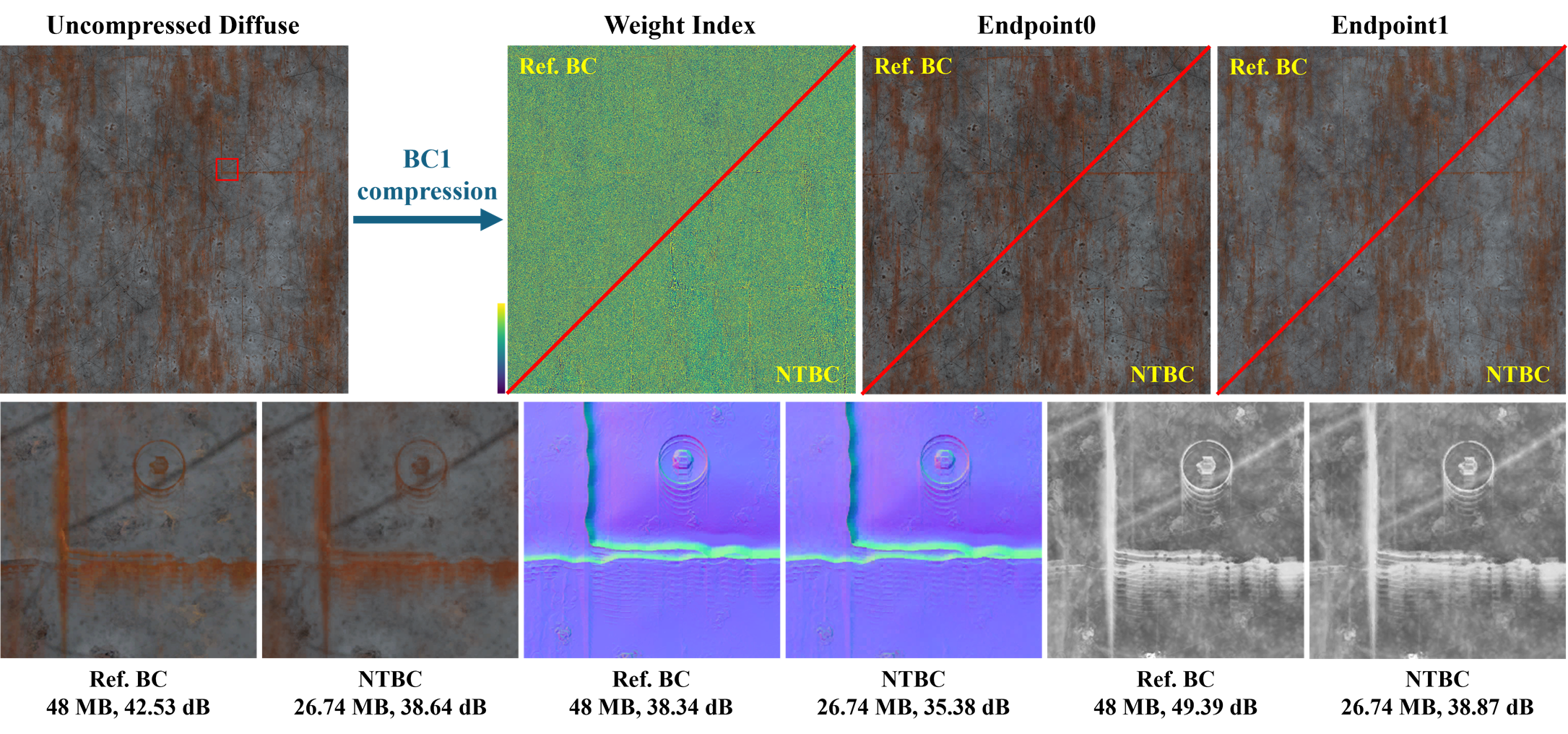}
  \caption{Neural Texture Block Compression (NTBC) encodes multiple textures in a single material in BC formats with reduced storage size while maintaining reasonable quality.
  Top row: block-compressed data in BC1 for diffuse. NTBC (bottom-right) produces the equivalent compressed data to the reference BC (Ref. BC, top-left).
  \textit{Weight Index} visualizes $2$-bit indices for each pixel with the viridis colormap where colors from black to yellow represent $0$ to $3$.
  Bottom row: zoom-ins of three textures in the \textit{MetalPlates013} material from~\cite{ambientCG} such as, from left to right, diffuse, normal, and roughness.
  }
  \label{fig:teaser}
}

\maketitle

\begin{abstract}
  Block compression is a widely used technique to compress textures in real-time graphics applications, offering a reduction in storage size.
  However, their storage efficiency is constrained by the fixed compression ratio, which substantially increases storage size when hundreds of high-quality textures are required.
  In this paper, we propose a novel block texture compression method with neural networks, Neural Texture Block Compression (NTBC).
  NTBC learns the mapping from uncompressed textures to block-compressed textures, which allows for significantly reduced storage costs without any change in the shaders.
  Our experiments show that NTBC can achieve reasonable-quality results with up to about $70\%$ less storage footprint, preserving real-time performance with a modest computational overhead at the texture loading phase in the graphics pipeline.

\end{abstract}

\section{Introduction}
The desire for more immersive experiences in games and virtual reality increases the demand for high visual fidelity in real-time graphics applications.
To meet this demand, textures play a crucial role in providing detailed and realistic surfaces through a lot of material properties such as diffuse color, normal maps, and other BRDF information.
However, each material property requires a high-resolution texture which consumes a large amount of storage and is often the bottleneck for the performance of graphics applications.
Block Compression (BC) is one of the most popular techniques to reduce the storage footprint of textures.
BC has different types of formats (BC1-BC7), which are supported by most modern GPUs and graphics APIs such as DirectX~\cite{DirectX}.
These formats offer the desirable random-access property with fixed-rate block compression, where each $4 \times 4$ texel block is compressed to a fixed number of bytes.
BC1 and BC4 compress each block to $8$ bytes, while other formats compress each block to $16$ bytes.
Therefore, a single $4$k texture compressed even with BC1 and BC4 requires $8$ MB of storage, which reaches the magnitude of gigabytes for a scene with hundreds of $4$k textures that are common in modern high-quality games.

Recent formats in BC such as BC6H and BC7 achieve higher-quality compression with a variety of modes and spatial partitioning patterns for each block.
More recently, variable-rate compression formats such as ASTC~\cite{ASTC} offer a well-balanced compression between quality and storage, using variable block sizes and flexible bit rates.
However, these formats require more expensive computations to find optimal configurations for each block.

In this work, we propose a novel neural BC approach, Neural Texture Block Compression (NTBC), that reduces the storage requirements of BC formats.
NTBC employs multi-layer perceptions (MLPs) to simultaneously encode block-compressed data of all textures in one material, achieving lower bit rates than the standard BC while producing the same block-compressed data format.

NTBC is meant to reduce the texture footprint on the disk. 
Network weights are stored in the disk which are loaded into the memory.
Then inference is executed to reconstruct block-compressed texture data which are copied to VRAM.
Therefore, it does not require any change in the shaders, which makes our method easier to adopt in the existing graphics pipelines.
We also utilize multi-resolution feature grids to encourage model optimization and compress them through quantization-aware training to reduce storage costs.
This paper focuses on the BC1 and BC4 formats, which are the simplest and most widely used for RGB and single-channel textures, respectively.

The main contributions of this work are as follows:
\begin{itemize}
    \item We introduce Neural Texture Block Compression (NTBC), a novel block compression based on MLPs optimized specifically for each material.
    \item We use multi-resolution feature grids compressed through quantization-aware training for better model optimization and storage efficiency.
    \item We demonstrate that NTBC reduces the storage cost of BC textures with minimal quality loss.
    \item Our experiments show that NTBC can encode multiple $4$k textures in one material within $10$ minutes and infer block-compressed data with a modest overhead on a single GPU.
\end{itemize}
\section{Related Work}
We propose a neural texture compression method following BC standards with compressed feature grids.
Thus, in this section, we begin by reviewing traditional BC and neural texture compression methods, and then we present a brief overview of quantization techniques for neural networks.

\subsection{Traditional Texture Compression}
The random access property of texture data is crucial for the efficient handling of compressed textures in real-time graphics pipelines on GPUs.
Delp and Mitchell~\cite{1094560} introduce block truncation coding (BTC) which is the first encoding method with fixed bits per block for greyscale images.
The method divides the image into non-overlapping $4 \times 4$ blocks and encodes them with two $8$-bit values and 1-bit indices to choose one of these two values.

Modern texture compression standards follow this BTC approach, including the S3 texture compression (S3TC) schemes~\cite{S3TC}.
S3TC is the first approach extending BTC to RGB images, which is later called DXTC and renamed to BC1 -- BC3 in DirectX~\cite{DirectX}.
BC1 encodes $4 \times 4$ blocks into $8$-byte structures.
This structure has two $16$-bit RGB565 endpoints and a set of $16$ $2$-bit indices to look up the corresponding texel color from a color palette.
The palette contains four colors with two additional colors linearly interpolated between the endpoints.
BC1 has another mode to support a $1$-bit alpha channel, but it is not our focus in this work.

BC4 is a specialized version of BC1 for single-channel images.
While utilizing the same 8-byte block size as BC1, BC4 employs two $8$-bit endpoints and a $3$-bit index per pixel to reference an 8-color palette.
This palette consists of the endpoints and either four or six linearly interpolated values, depending on whether the first endpoint numerically exceeds the second.
In cases with fewer interpolated values, special values $0$ and $1$ are incorporated.
The expanded palette and increased endpoint precision enable BC4 to achieve higher quality than BC1 for single-channel cases.
BC2, BC3, and BC5 are simply combinations of BC1 and BC4, so we omit their further descriptions.
This work focuses on BC1 and BC4 to encode RGB and single-channel textures, respectively.

All these methods presented so far are designed to encode all the pixels in a block as a single entity.
On the other hand, the idea of partitioning a block into multiple entities and encoding them separately is proposed~\cite{ETC2}.
BC7 further extends this idea by supporting a variety of partition shapes and encoding each partition with different endpoints and indices.
More recently, ASTC~\cite{ASTC} has been introduced as a more flexible and efficient approach.
ASTC computes the optimal partitioning using a hash function and encodes variable-sized blocks to different numbers of bytes.
These modern complex methods are not our focus in this paper, but extending our method to support them is an interesting future work.

\subsection{Neural Texture Compression}
Neural network-based image compression methods commonly use a \emph{multi-resolution feature grid} that stores latent embeddings of the input images in each grid cell~\cite{instantNGP, takikawa2023compact}.
For texture compression that requires random access queries, Vaidyanathan et al.~\cite{ntc2023} propose a neural compression method specifically designed for textures.
They use a specialized architecture to encode multiple textures and their mipmap chains together with a small MLP and compressed representation of tailored feature grids.
Compact NGP~\cite{takikawa2023compact} is the method of compressing feature grids by indexing the spatial hash table using the learned indexing codebook for collision detection.
It is not specialized for texture compression, but it similarly has the random-access property which enables its application to texture compression.
Block-compressed features (BCf)~\cite{BCf} is the novel approach using BC6 to compress learned neural texture features and decompress them in a real-time renderer.
It uses BC as the means of compressing neural features instead of learning the encoding of block-compressed data.
Inspired by these neural texture compression methods, our method also encodes multiple textures with compressed feature grids decoded by small MLPs.
However, unlike them, we focus on block compression itself and propose a method to encode block-compressed data that can be easily integrated into existing graphics pipelines.

Additionally, Pratapa et al.~\cite{TexNN} uses a neural network (NN) for texture compression from a different perspective.
They propose TexNN which replaces the expensive search step of the optimal configuration such as partitioning and endpoints format in the recent texture compression methods like BC7 and ASTC with a neural network.

\subsection{Neural Network Quantization} \label{sec:nnq}
Quantization is a technique to reduce the network size and its inference costs by converting network weights and activations from high-precision floating points to low-precision integers.
There are two common approaches to quantizing NNs: Post-Training Quantization (PTQ) and Quantization-Aware Training (QAT)~\cite{SurveyNNQ,WPNNQ}.
PTQ quantizes the weights of a pre-trained network without any fine-tuning, which is very fast but often comes with a significant loss in accuracy.
On the other hand, QAT fine-tunes the network with quantized parameters so that the network can recover the accuracy loss caused by quantization.

The standard operations used for quantization are based on uniform affine transformations between a high-precision real value $r$ to a low-precision integer $q$ as follows~\cite{app12157829}:
\begin{align}
    q &= \text{clamp} \left( \lfloor \frac{r}{s} \rceil + z; n, p \right), \label{eq:quantize} \\
    r &= s \cdot ( q - z ), \label{eq:dequantize}
\end{align}
where $\lfloor \cdot \rceil$ denotes the half-way rounding function, $s$ is the scale factor which is a positive real value, and $z$ is the zero point which is an integer value that ensures the real zero is representable by an integer value in the quantized domain.
$n$ and $p$ are the minimum and maximum representable values, respectively.
Given that $q$ is an unsigned integer, $n = 0$ and $p = 2^b - 1$, where $b$ is the target bit width.
We refer to the operators that perform these transformations to emulate quantization during training as \textit{quantizers} which can be represented using Eq.~\ref{eq:quantize} and Eq.~\ref{eq:dequantize} as the following function:
\begin{equation} \label{eq:quantizer}
    Q(w) = s \cdot \left( \text{clamp} \left( \lfloor \frac{w}{s} \rceil + z; n, p \right) - z \right),
\end{equation}
where $w$ represents the real-valued weights of the network.

The choice of the scaling factor $s$ is crucial for the quantization accuracy.
It divides the range of the real value into $2^b$ intervals:
\begin{equation} \label{eq:scaling_factor}
    s = \frac{\beta - \alpha}{2^b - 1},
\end{equation}
where $[\alpha, \beta]$ is the range of the real value to be quantized.
Therefore, to determine the optimal scaling factor, the range of the real value should be estimated.
Using the minimum and maximum values of the weights is a common approach to estimating the range, which is often referred to as \textit{asymmetric quantization} because the range is not necessarily symmetric concerning the origin.
And, in practice, the zero point $z$ is computed as
\begin{equation} \label{eq:zero_point}
    z = \lfloor \alpha \cdot s \rceil.
\end{equation}
Note that the straight-through estimator (STE)~\cite{STE} is commonly used during training to avoid extremely sparse gradients due to the rounding function so that $\nabla_x \lfloor x \rceil = 1$.
Considering Eq.~\ref{eq:quantizer}, the backward pass of QAT only depends on the clamping function:
\begin{equation}
    \nabla_w Q(w) =
    \begin{cases}
        w, & \text{if } w \text{ is not clamped}, \\
        0, & \text{otherwise}.
    \end{cases}
\end{equation}

To quantize the feature grids, Vaidyanathan et al.~\cite{ntc2023} simulate quantization by adding uniform noise to the features in grids and using a fixed quantization range instead of computing the optimal range $[\alpha, \beta]$ during training.
Instead, we quantize the feature grids by applying QAT of NNs to the grids directly, which is more principled and allows for better quantization accuracy.
\section{Method} \label{sec:method}
Colors on the palette $c_n$ are linearly interpolated between two endpoints for each block $\mathbf{e_0} \in \mathbb{R}^3, \mathbf{e_1} \in \mathbb{R}^3$ for BC1 (or $e_0 \in \mathbb{R}^1, e_1 \in \mathbb{R}^1$ for BC4) with weights $w_n$ with the following equation:
\begin{equation} \label{eq:palette}
    \mathbf{c_n} = (1 - w_n) \cdot \mathbf{e_0} + w_n \cdot \mathbf{e_1},
\end{equation}
where $n$ is a $2$-bit index per pixel ($0 \leq n \leq 3$) for BC1 and a $3$-bit index per pixel ($0 \leq n \leq 7$) for BC4.
Weights $w_n$ for BC1 can be represented as $w_n = \frac{n}{3}$.
However, depending on whether the first endpoint numerically exceeds the second, the palette of BC4 contains either four or six linearly interpolated values, with special values 0 and 1 in the case with fewer interpolated values.
Therefore, if $e_0 > e_1$, weights $w_n$ for BC4 can be represented as $w_n = \frac{n}{7}$.
On the other hand, if $e_0 \leq e_1$, we can represent them as the following equation:
\begin{equation}
    \label{eq:weightBC4}
    w_n =
        \begin{cases}
            \frac{e_0}{e_0 - e_1} & n = 0, \\
            \frac{n-1}{5} & 1 \leq n \leq 6, \\
            \frac{1 - e_0}{e_1 - e_0} & n = 7.
        \end{cases}
\end{equation}
With Eq.~\ref{eq:palette} and the first and third cases of Eq.~\ref{eq:weightBC4}, we achieve $c_n = 0$ and $c_n = 1$, respectively.

\begin{figure}
    \centering
    \includegraphics[width=\columnwidth]{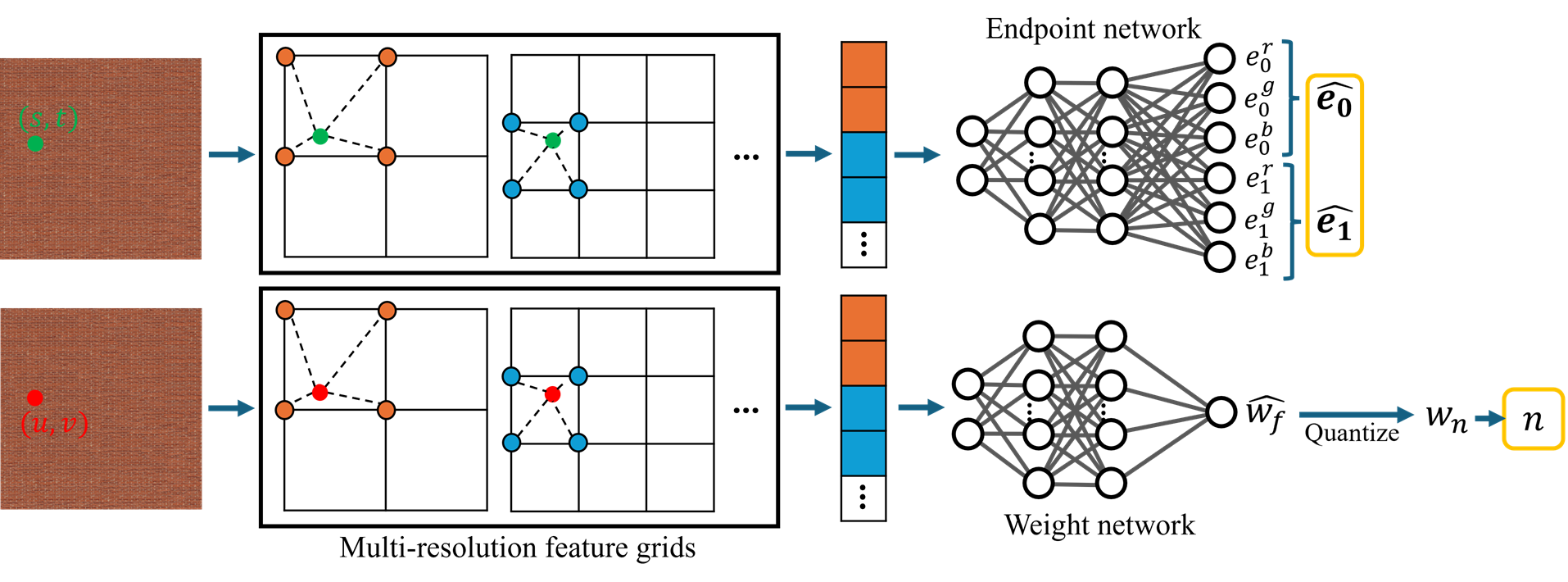}
    \caption{Naive approach. Two MLPs infer weights and endpoints.
    The weight network is trained to output continuous weights $\hat{w_f}$, and the endpoint network is later fine-tuned with quantized weights $w_n$.
    Values in yellow squares form a compressed block.}
    \label{fig:naiveNet}
\end{figure}

\begin{figure*}
    \centering
    \begin{subfigure}[b]{0.50\textwidth}
        \centering
        \includegraphics[width=\textwidth]{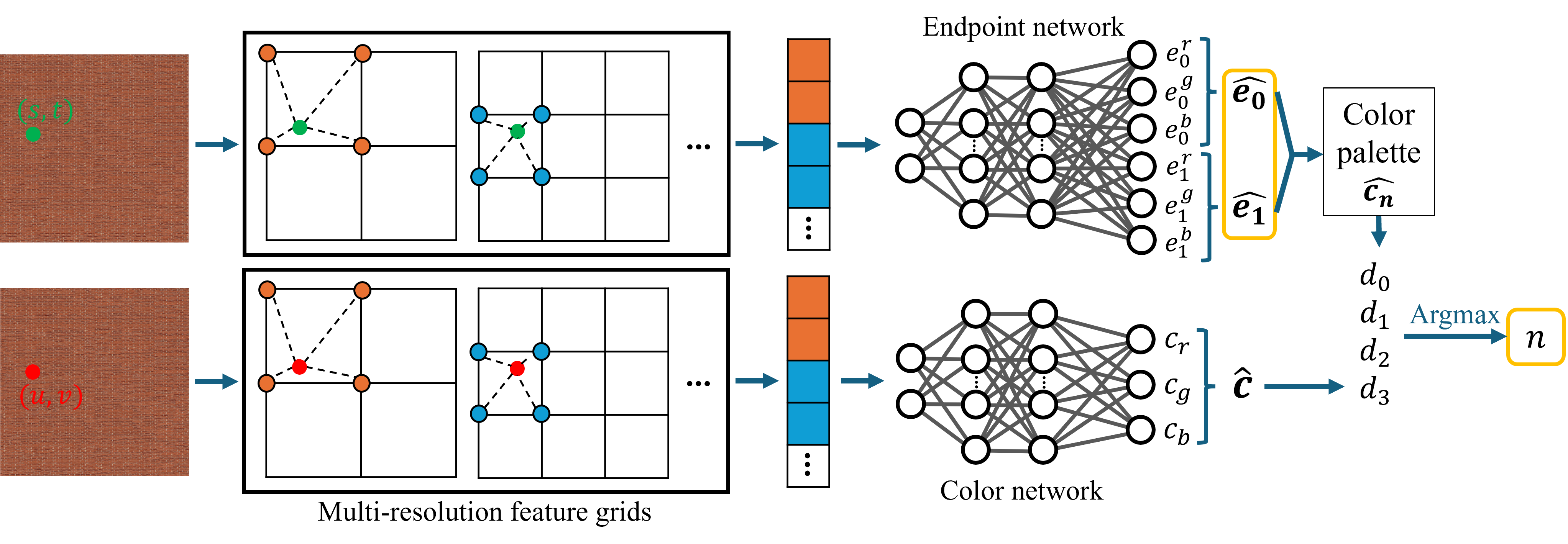}
        \caption{\small{Inference procedure}}
        \label{fig:NTBC_predict}
    \end{subfigure}
    \hfill
    \begin{subfigure}[b]{0.48\textwidth}
        \centering
        \includegraphics[width=\textwidth]{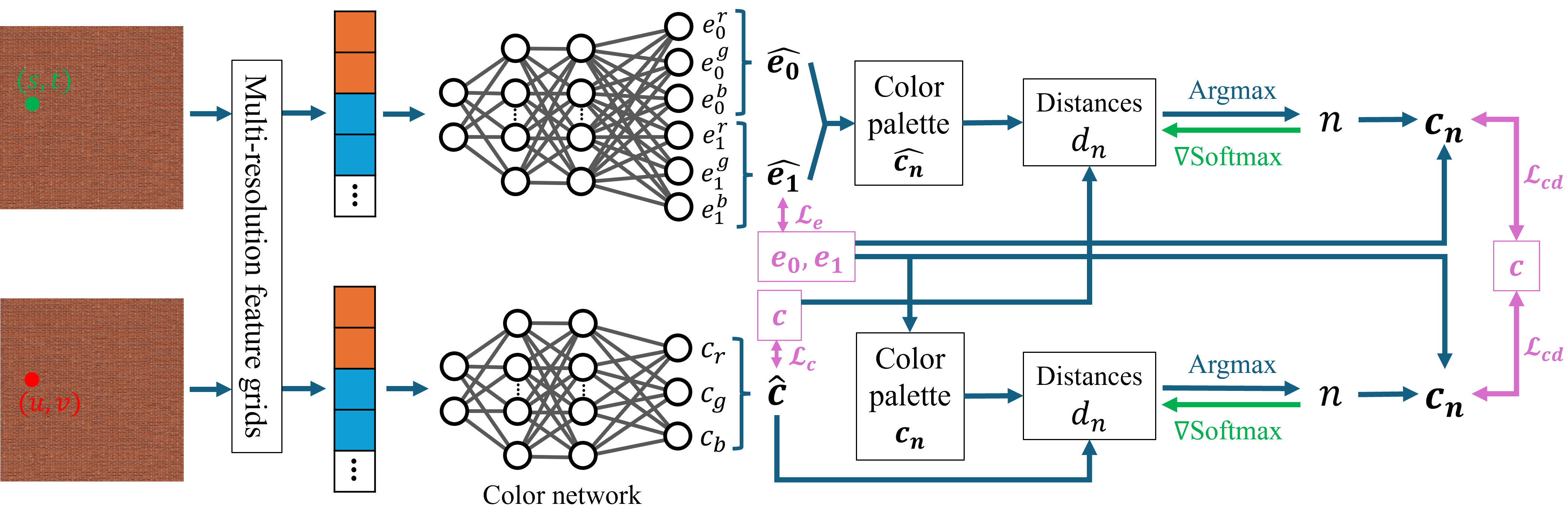}
        \caption{\small{Training procedure}}
        \label{fig:NTBC_train}
    \end{subfigure}
    \caption{Neural Texture Block Compression. Two MLPs infer endpoints and uncompressed colors.
    (a) In the inference procedure, we compute weight indices $n$ using predicted endpoints and colors. A resulting compressed block is indicated by yellow squares.
    (b) In the training procedure, two networks are trained with reference endpoints ($\mathcal{L}_e$) and uncompressed colors ($\mathcal{L}_c$), and additional errors of decoded colors ($\mathcal{L}_{cd}$).
    Purple indicates reference values and their losses. For optimization, we propagate the gradients through the argmax operation as if an index $n$ was computed with the softmax operation.}
    \label{fig:NTBC}
\end{figure*}

Our goal is to construct a neural model encoding block-compressed data in BC1 and BC4 formats, and these $\mathbf{e_0}, \mathbf{e_1}$, and $n$ (i.e $w_n$) consist of block-compressed data.
Therefore, the straightforward way to encode them using NNs is to directly optimize them through differentiation of Eq.~\ref{eq:palette}.
Fig.~\ref{fig:naiveNet} illustrates the architecture of this \textit{naive approach}, which we will explain further in Sec.~\ref{sec:naive}.
However, the resulting block-compressed data with the naive approach show only limited-quality results in our experiments.
This is because weights are high-frequency and not spatially correlated, which makes it hard for MLPs to be directly optimized~\cite{pmlr-v97-rahaman19a}.
To solve this problem, our method, NTBC, predicts the original uncompressed colors instead of weights as shown in Fig.~\ref{fig:NTBC}.
Uncompressed colors are lower-frequency and more spatially correlated than weights, which is easier for MLPs to learn.
We will further describe our method in Sec.~\ref{sec:ntbc}.

Material properties are usually represented by multiple textures.
Traditional block compression approaches can only compress each texture individually which is memory-intensive for materials (e.g. $8$ MB for a single $4$k texture in BC1 or BC4 formats).
On the other hand, NTBC encodes multiple textures at once by being optimized for each material, assuming a significant correlation across different textures as done by Vaidyanatha et al.~\cite{ntc2023}.
In this section, we first describe the naive approach and then extend it to NTBC, considering BC1 with a single texture for brevity.
The methods can be easily extended to multiple textures by modifying the number of output nodes in each network.
They can be also extended to BC4 by modifying the color space from $\mathbb{R}^3$ to $\mathbb{R}^1$ and the weight index computation considering Eq.~\ref{eq:weightBC4}.

\subsection{Naive Approach} \label{sec:naive}
As described in Fig.~\ref{fig:naiveNet}, we prepare two different MLPs to predict endpoints and weights, referred to as the \emph{endpoint network} and the \emph{weight network}, respectively.
The endpoint network takes the 2D normalized block indices, $s, t \in [0, 1]$, as inputs and predicts endpoints $\mathbf{\hat{e_0}}, \mathbf{\hat{e_1}}$ for each block.
The weight network uses the 2D texture coordinates, $u, v \in [0, 1]$, as inputs, infers floating-point weights $\hat{w_f}$ for each texel, and then, quantizes $\hat{w_f}$ to $w_n$ to construct block-compressed data.
These 2D inputs are encoded with multi-resolution feature grids which excel at representing spatially-varying features.

The naive approach is trained to minimize the discrepancy between reference uncompressed colors and decoded colors derived from $\mathbf{\hat{e_0}}, \mathbf{\hat{e_1}}$, and $w_n$ with Eq.~\ref{eq:palette}.
To enhance training efficiency and optimize for discrete weights $w_n$, a two-stage training procedure is employed:
During the initial $80\%$ of training iterations, an $\mathcal{L}_2$ loss function is applied to decoded colors generated from $\mathbf{\hat{e_0}}, \mathbf{\hat{e_1}}$, and floating-point weights $\hat{w_f}$ with Eq.~\ref{eq:palette}.
Subsequently, the endpoint network is fine-tuned for the remaining iterations using quantized weights $w_n$, while the weights of the weight network are frozen.

\subsection{Neural Texture Block Compression} \label{sec:ntbc}
Our method, NTBC, is a novel method that encodes block-compressed data in BC1 and BC4 formats using NNs.
As shown in Fig.~\ref{fig:NTBC}, NTBC employs two networks similar to the naive approach, but instead of the weight network, the \textit{color network} predicts the original uncompressed colors.
Uncompressed colors are lower-frequency and more spatially correlated than weights, which is easier for MLPs to learn.
Inputs to the color network are 2D texture coordinates encoded with multi-resolution feature grids, and it infers uncompressed colors $\mathbf{\hat{c}}$.

To construct block-compressed data from the predicted $\mathbf{\hat{e_0}}, \mathbf{\hat{e_1}}$, and $\mathbf{\hat{c}}$, we first compute colors on the palette $\mathbf{\hat{c_n}}$ from $\mathbf{\hat{e_0}}$ and $\mathbf{\hat{e_1}}$ using Eq.~\ref{eq:palette}, as shown in Fig.~\ref{fig:NTBC_predict}.
Second, we calculate distances $d_n$ between $\mathbf{\hat{c_n}}$ and $\mathbf{\hat{c}}$ in the color space.
Then, a $2$-bit index $n$ for each texel is computed from $d_n$.
For simplicity, distances $d_n$ are computed as negative values which allows for using the argmax operation to achieve $n$:
\begin{align}
    d_n &=
    \begin{cases}
        - \lVert \mathbf{c} - \mathbf{\hat{c_n}} \rVert &\text{for endpoint network}, \\
        - \lVert \mathbf{\hat{c}} - \mathbf{c_n} \rVert &\text{for color network},
    \end{cases} \\ \label{eq:dist}
    n &= \text{argmax} (d_n).
\end{align}

The endpoint and color networks are trained independently using the combined loss to facilitate the optimizations for both colors and weight indices.
Fig.~\ref{fig:NTBC_train} illustrates the training procedure.
First, the endpoint and color networks learn the optimal endpoints and uncompressed colors with an $\mathcal{L}_2$ loss for reference endpoints $\mathbf{e_0}, \mathbf{e_1}$ ($\mathcal{L}_e$) and reference uncompressed colors $\mathbf{c}$ ($\mathcal{L}_c$), respectively.
Reference endpoints $\mathbf{e_0}, \mathbf{e_1}$ are obtained from block-compressed data computed with Compressonator~\cite{Compressonator}.
Also, for both networks, we use the additional $\mathcal{L}_2$ loss for decoded colors derived from $\mathbf{e_0}, \mathbf{e_1}$ and reconstructed $2$-bit indices $n$ ($\mathcal{L}_{cd}$) to consider an error from the weight index computation.
Therefore, the combined training loss we use for each network is:
\begin{align} \label{eq:lossNTBC_e}
    \mathcal{L}_{endpoint} &= \mathcal{L}_e + \mathcal{L}_{cd}, \\
    \mathcal{L}_{color} &= \mathcal{L}_c + \mathcal{L}_{cd}. \label{eq:lossNTBC_c}
\end{align}
As shown in Fig.~\ref{fig:NTBC_train}, to compute a $2$-bit index $n$ for each texel, we use predicted endpoints $\mathbf{\hat{e_0}}, \mathbf{\hat{e_1}}$ and reference colors $\mathbf{c}$ for the endpoint network while reference endpoints $\mathbf{e_0}, \mathbf{e_1}$ and predicted uncompressed colors $\mathbf{\hat{c}}$ are used for the color network.
With this weight index $n$ and reference endpoints $\mathbf{e_0}, \mathbf{e_1}$, the final decoded colors are computed with Eq.~\ref{eq:palette}.

However, the argmax operation is not differentiable, so we cannot directly optimize the networks using the loss $\mathcal{L}_{cd}$ in Eq.~\ref{eq:lossNTBC_e} and Eq.~\ref{eq:lossNTBC_c}.
To solve this problem, we use STE~\cite{STE} to backpropagate the gradients through the argmax operation.
In the forward pass, we take the argmax operation of $d_n$ to obtain a weight index $n$, while in the backward pass, we propagate the gradients of the softmax of $d_n$ weighted by the corresponding weight $w_n = \frac{n}{3}$.
In other words, the gradients in the backward pass can be computed as if we took the expectation of the weights $w_n$ with the normalized probability computed with the softmax operation.
Further details of this gradient computation are described in our supplemental document.

The softmax operation normalizes the distances $d_n$ to the probability distribution as the following equation:
\begin{equation} \label{eq:softmax}
    \sigma(d_n) = \frac{\exp(d_n / T)}{\sum_i \exp(d_i / T)},
\end{equation}
where $T$ is the temperature parameter controlling the sharpness of the distribution.
Larger $T$ makes the probability distribution more uniform while smaller $T$ makes it more concentrated on the largest $d_n$.
Our experiments use the small $T$ to make the gradients shaper and the training more stable.

\subsection{Grids Quantization} \label{sec:quantization}
Our goal is to encode block-compression data of multiple textures with a reduced storage size.
To achieve this, we quantize the feature grids through QAT described in Sec.~\ref{sec:nnq}.
We empirically found that the $8$-bit quantization can achieve equivalent-quality results to the half-precision floating points, processed with the following procedure:
We first train an NTBC model as described in Sec.~\ref{sec:ntbc} with the half-precision floating points.
Then, for $10\%$ more steps, we quantize only the feature grids into $8$-bit discrete values using QAT with Eq.~\ref{eq:quantizer} and fine-tune MLPs with the quantized feature grids.
Once the model is trained, we store the feature grids as $8$-bit integers on the disk which are loaded into the memory.
Then, the inference is executed in the half-precision floating points by dequantizing them with Eq.~\ref{eq:dequantize}.

To apply QAT to the multi-resolution feature grids, we conduct asymmetric quantization for each level of the grids.
Therefore, we compute the ranges of the real value $[\alpha, \beta]$ by taking the minimum and maximum values in each level of the grids.
Then, the scaling factor $s$ and zero-point $z$ are computed with Eq.~\ref{eq:scaling_factor} and Eq.~\ref{eq:zero_point} and are dynamically updated in each training iteration.
Using these parameters for each level with Eq.~\ref{eq:quantizer}, we emulate the $8$-bit quantization of the multi-resolution feature grids in the training procedure.

\subsection{Model Configurations} \label{sec:config}
We use small MLPs for all networks having three hidden layers with $64$ neurons each with half-precision floating points.
A sigmoid activation function is applied to all output layers to predict the values in the range of $[0, 1]$.
Empirical evaluation of various activation functions for three hidden layers indicated that a \emph{selu} activation function~\cite{selu} is the most effective.
Multi-resolution feature grids have different configurations for 2D block indices and texture coordinates.
The 2D block indices are represented using $7$ levels with the finest resolution of $1,024$, while $8$ levels with the finest resolution of $2,048$ are employed for the 2D texture coordinates.
All grids comprise 2D features per level and a coarsest resolution of $16$.
These grid configurations result in $14$-dimensional inputs for the endpoint network and $16$-dimensional inputs for the weight and color networks.
Reducing the model size, we compress these feature grids through QAT into 8-bit integers as described in Sec.~\ref{sec:quantization}.

We jointly optimize multi-resolution feature grids and MLPs using gradient descent with the Adam optimizer~\cite{Adam}, where we set $\beta_1 = 0.9$, $\beta_2 = 0.999$, and $\epsilon = 10^{-15}$.
We use a high initial learning rate of $0.01$ for feature grids and a low initial learning rate of $0.005$ for MLPs and apply cosine annealing~\cite{SGDR} with warm-up iterations of $10$ to lower the learning rate to $0$ at the end of training.
The weights of MLPs are initialized with He initialization procedure~\cite{HeInit} and stored in half-precision floating-point formats.
And, feature grids are initialized with the uniform distribution in the range of $[-10^{-4}, 10^{-4}]$ and stored in quantized 8-bit integers.
Also, for the softmax computation in Eq.~\ref{eq:softmax}, we use the small temperature $T = 0.01$ in all our experiments.

\begin{figure*}
    \centering
    \includegraphics[width=\textwidth]{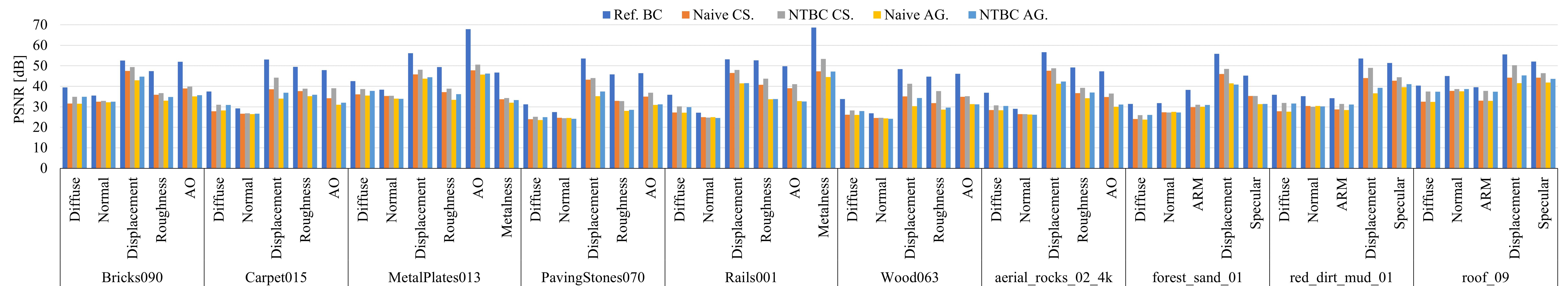}
    \caption{Quantitative comparison for all materials in our dataset for different methods.
    Materials are retrieved from~\cite{ambientCG} and~\cite{polyhaven}.
    \textit{CS.} and \textit{AG.} are short for \textit{conservative} and \textit{aggressive}, respectively.
    Diffuse, Normal, and ARM are RGB textures compressed as BC1, while others are single-channel textures compressed as BC4.}
    \label{fig:quantitative}
\end{figure*}

\section{Results} \label{sec:results}
We implemented the naive approach and NTBC from scratch using C++ and HIP~\cite{hip} for GPU programming language to train the models and run inference on GPUs.
Inspired by the previous work from Müller et al.~\cite{NRC}, our implementation is to fuse operations from all layers into a single kernel to minimize the overhead of memory access and kernel launch.
We used the same model parameters and training configurations as much as possible both for the naive approach and NTBC (Sec.~\ref{sec:config}) for a fair comparison.
All the evaluations in this paper are performed on a single AMD Radeon\textsuperscript{\texttrademark} RX $7900$ XT GPU on a Windows machine.

\subsection{Evaluation Methods} \label{sec:evalMethod}
Our proposed method, NTBC, encodes block-compressed data in BC1 and BC4 formats in this research.
To the best of our knowledge, no existing method compresses textures using NNs in BC formats.
Therefore, in this paper, we evaluate NTBC compared to the reference BC computed with Compressonator~\cite{Compressonator} and the naive approach introduced in Sec.~\ref{sec:naive}.
To compress RGB textures in BC1 using Compressonator, we used two refine steps for the compression setting.

The naive approach and NTBC encode multiple textures in a material at once.
A material usually contains both RGB and single-channel textures to represent variable material properties, which are compressed in the different BC formats, BC1 and BC4, respectively.
These two formats have different numbers of channels and colors in a palette, which makes it inefficient to compress them together.
Then, in this paper, we evaluate two different approaches to compressing RGB and single-channel textures separately and together: \textit{conservative} and \textit{aggressive}.
In the following descriptions, we refer to the numbers of RGB and single-channel textures in a material as $N_{RGB}$ and $N_{SC}$, respectively.

\paragraph*{Conservative approach.}
This approach compresses RGB and single-channel textures separately using two distinct models.
For RGB textures, the model is trained with BC1 compression, employing $6 \cdot N_{RGB}$, $N_{RGB}$, and $3 \cdot N_{RGB}$ output nodes for the endpoint, weight, and color networks, respectively.
On the other hand, single-channel texture compression relies on a BC4-trained model with $2 \cdot N_{SC}$ and $N_{SC}$ output nodes for the endpoint and weight/color networks, respectively.
While straightforward, this approach suffers from inefficiency due to the storage overhead of maintaining two separate models.

\paragraph*{Aggressive approach.}
This approach utilizes a single model to compress both RGB and single-channel textures to improve storage efficiency.
This model is trained jointly for BC1 and BC4 by incorporating dummy nodes in computing distances $d_n$.
Specifically, while eight distance nodes ($d_0$ to $d_7$) are used for BC4, only the first four are active during BC1 training, with the remaining four serving as dummies.
Gradients for dummy nodes are zero during backpropagation.
The numbers of output nodes are $6 \cdot N_{RGB} + 2 \cdot N_{SC}$ for the endpoint network, $N_{RGB} + N_{SC}$ for the weight network, and $3 \cdot N_{RGB} + N_{SC}$ for the color network.
This approach reduces the storage footprint compared to the conservative approach but increases training complexity due to the need to learn the characteristics of both compression formats.

\subsection{Texture Dataset} \label{sec:dataset}
We collected $10$ different materials from ambientCG~\cite{ambientCG} and Poly Haven~\cite{polyhaven} to evaluate different block compression methods described in Sec.~\ref{sec:evalMethod}.
Materials contain different numbers of RGB and single-channel textures, and their contents represent varying characteristics such as noisy, smooth, and high-frequency patterns.
Resolutions of all textures in the materials are $4096^2$.
More details for this dataset can be found in our supplemental document.

\subsection{Evaluation} \label{sec:comparison}
\paragraph*{Quantitative results.}
We evaluate our method quantitatively using the PSNR and SSIM~\cite{SSIM} metrics computed against uncompressed textures, and higher values indicate better quality.
The naive approach and NTBC are trained for $20$k iterations in all the experiments with all textures in each material.

Fig.~\ref{fig:quantitative} shows PSNR metrics of the reference BC, the naive approach, and NTBC for all the textures in our dataset.
NTBC achieves higher PSNR than the naive approach for almost all textures, especially the conservative approach of NTBC shows the best results among all NN-based methods.
However, for normal textures, the naive approach has similar results to NTBC, which indicates that NTBC is more effective for textures with high-frequency patterns.
Also, comparing the conservative and aggressive approaches, the conservative one shows a significant improvement for single-channel textures, while it shows equivalent or a bit better results for RGB textures.
It indicates that the aggressive approach is more sensitive to texture characteristics across all the textures in the material, so we recommend users select the proper approach depending on the use cases to achieve the best results.
More detailed numbers for PSNR can be found in our supplemental document.

\begin{figure}
    \centering
    \includegraphics[width=0.9\columnwidth]{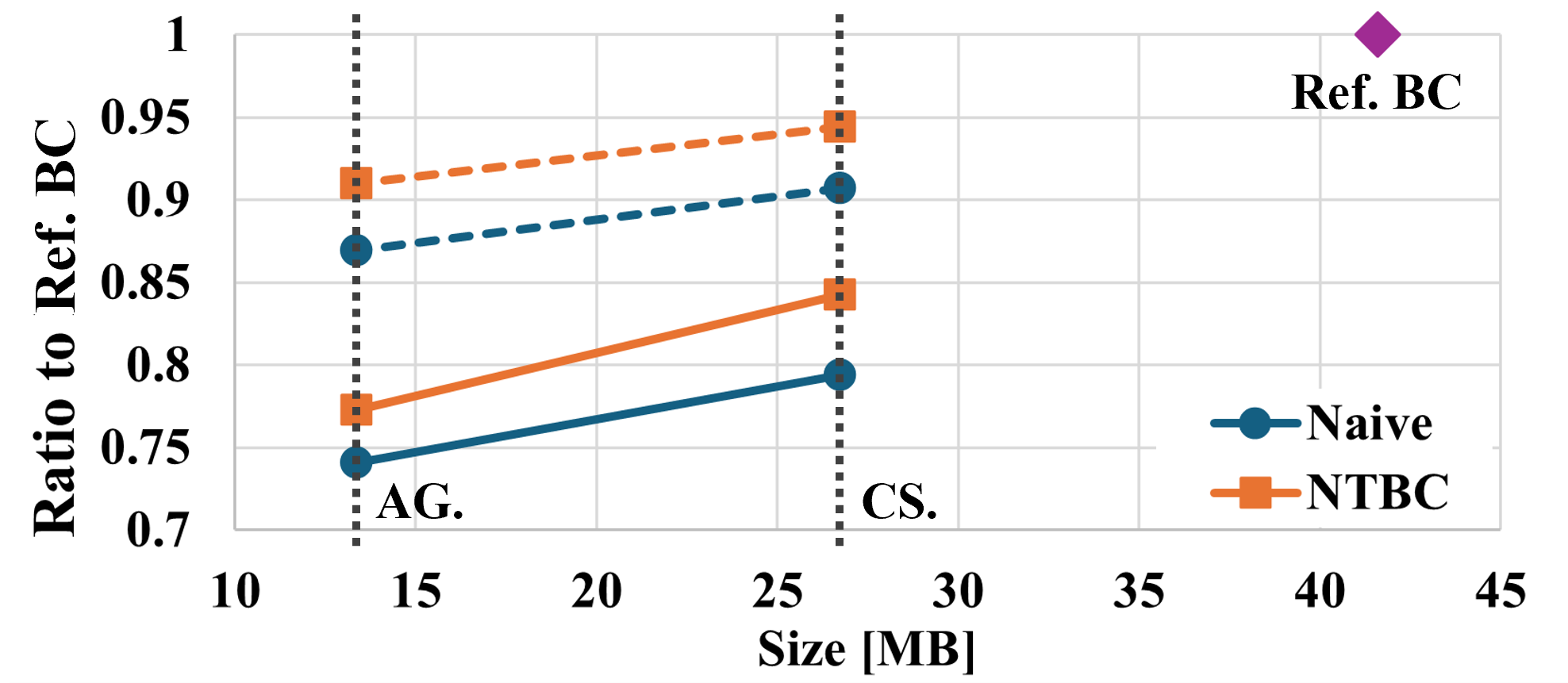}
    \caption{Ratio of average PSNR and SSIM of the naive approach and NTBC to the reference BC vs. their storage costs.
    The solid lines show the average PSNR ratio while the dashed lines show the average SSIM ratio.
    \textit{AG.} and \textit{CS.} are short for \textit{aggressive} and \textit{conservative}, respectively.}
    \label{fig:average}
\end{figure}

Fig.~\ref{fig:average} visually shows the ratio of average PSNR and SSIM of the naive approach and NTBC to the reference BC over their storage costs, which are averaged over all the materials in the dataset.
It illustrates the tradeoff between the storage cost and the quality of the compressed textures.
The conservative approach consumes $26.74$ MB of the storage which is about $40\%$ of BC with Compressonator ($41.6$ MB) while maintaining the quality with a reasonable degradation of about $15\%$ in PSNR and $5\%$ in SSIM.
And, more aggressively, the storage cost is reduced to $13.37$ MB by the aggressive one with a bit more quality loss of about $20\%$ in PSNR and $10\%$ in SSIM.
Thus, users can choose the appropriate approach of NTBC depending on the storage budget and the quality requirement.
Additionally, because NTBC is a simple replacement for the existing BC methods, the standard BC can be also used for the textures that NTBC cannot compress well.

\begin{figure*}
    \centering
    \setlength{\tabcolsep}{0\linewidth}
    \renewcommand{\arraystretch}{1.0}
    \begin{tabular}{c@{\hspace{0.004\linewidth}} c@{\hspace{0.002\linewidth}} |@{\hspace{0.002\linewidth}} c@{\hspace{0.002\linewidth}} c@{\hspace{0.002\linewidth}} |@{\hspace{0.002\linewidth}} c@{\hspace{0.002\linewidth}} c@{\hspace{0.002\linewidth}} |@{\hspace{0.002\linewidth}} c}
        & & \multicolumn{2}{c@{\hspace{0.002\linewidth}}|@{\hspace{0.002\linewidth}}}{ Aggressive } & \multicolumn{2}{c@{\hspace{0.002\linewidth}}|@{\hspace{0.002\linewidth}}}{ Conservative } & \\

        & & Naive & NTBC & Naive & NTBC & Ref. BC \\


        & Size & \multicolumn{2}{c@{\hspace{0.002\linewidth}}|@{\hspace{0.002\linewidth}}}{ 13.37 MB } & \multicolumn{2}{c@{\hspace{0.002\linewidth}}|@{\hspace{0.002\linewidth}}}{ 26.74 MB } & 40 MB \\

        \midrule

        \raisebox{0.08\linewidth}{\rotatebox[origin = c]{90}{Diffuse}} &
        \begin{overpic}[width=0.15\linewidth]{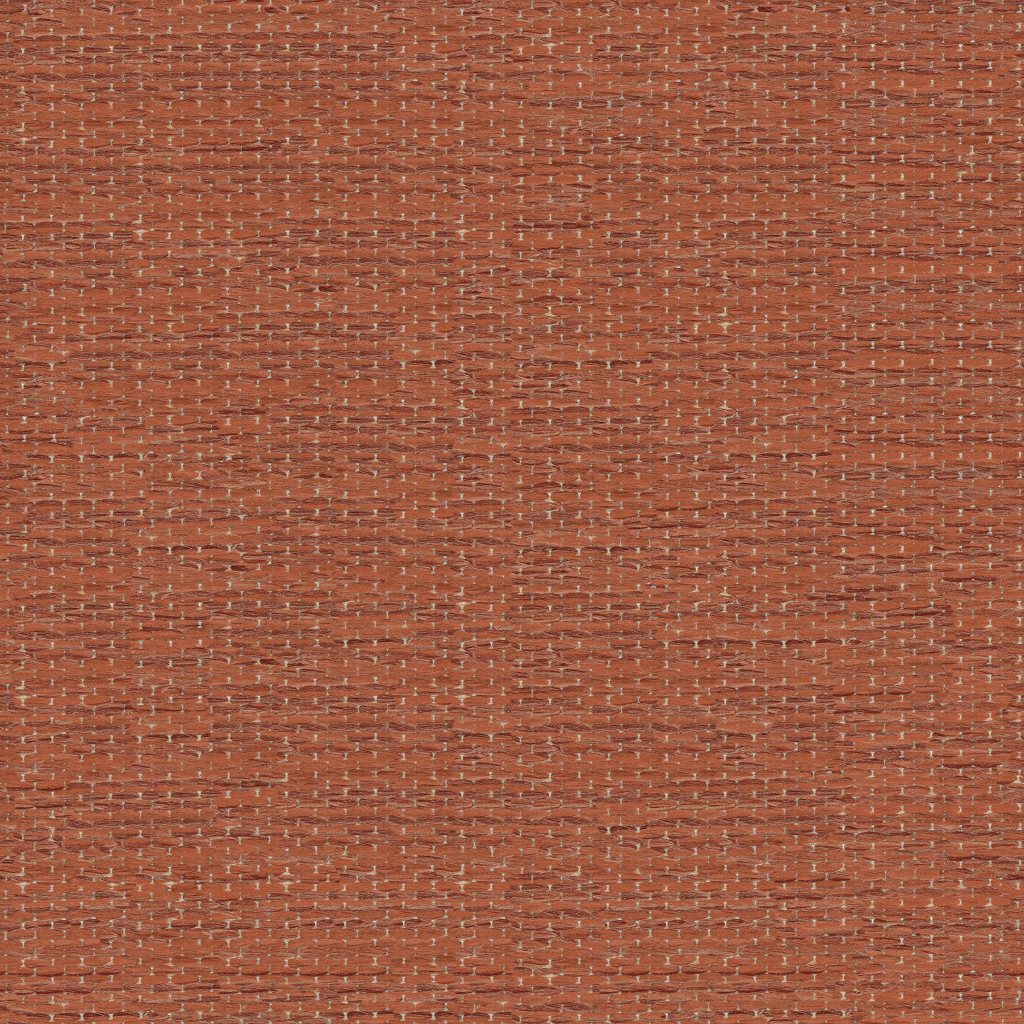}
            \put(795,580){\linethickness{0.2mm}\color{red}\polygon(0,0)(50,0)(50,50)(0,50)}
        \end{overpic} &
        \includegraphics[width=0.15\linewidth]{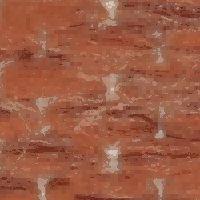} &
        \includegraphics[width=0.15\linewidth]{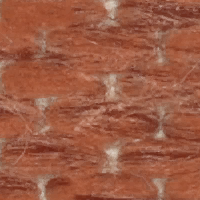} &
        \includegraphics[width=0.15\linewidth]{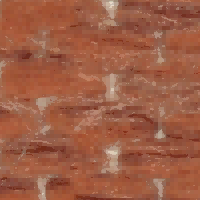} &
        \includegraphics[width=0.15\linewidth]{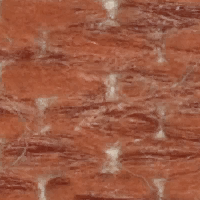} &
        \includegraphics[width=0.15\linewidth]{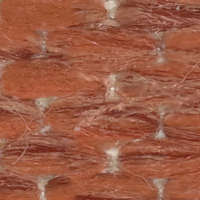} \\


        & PSNR & 28.21 dB & 30.90 dB & 27.81 dB & \textbf{30.98 dB} & 37.44 dB \\


        \raisebox{0.08\linewidth}{\rotatebox[origin = c]{90}{Displacement}} &
        \begin{overpic}[width=0.15\linewidth]{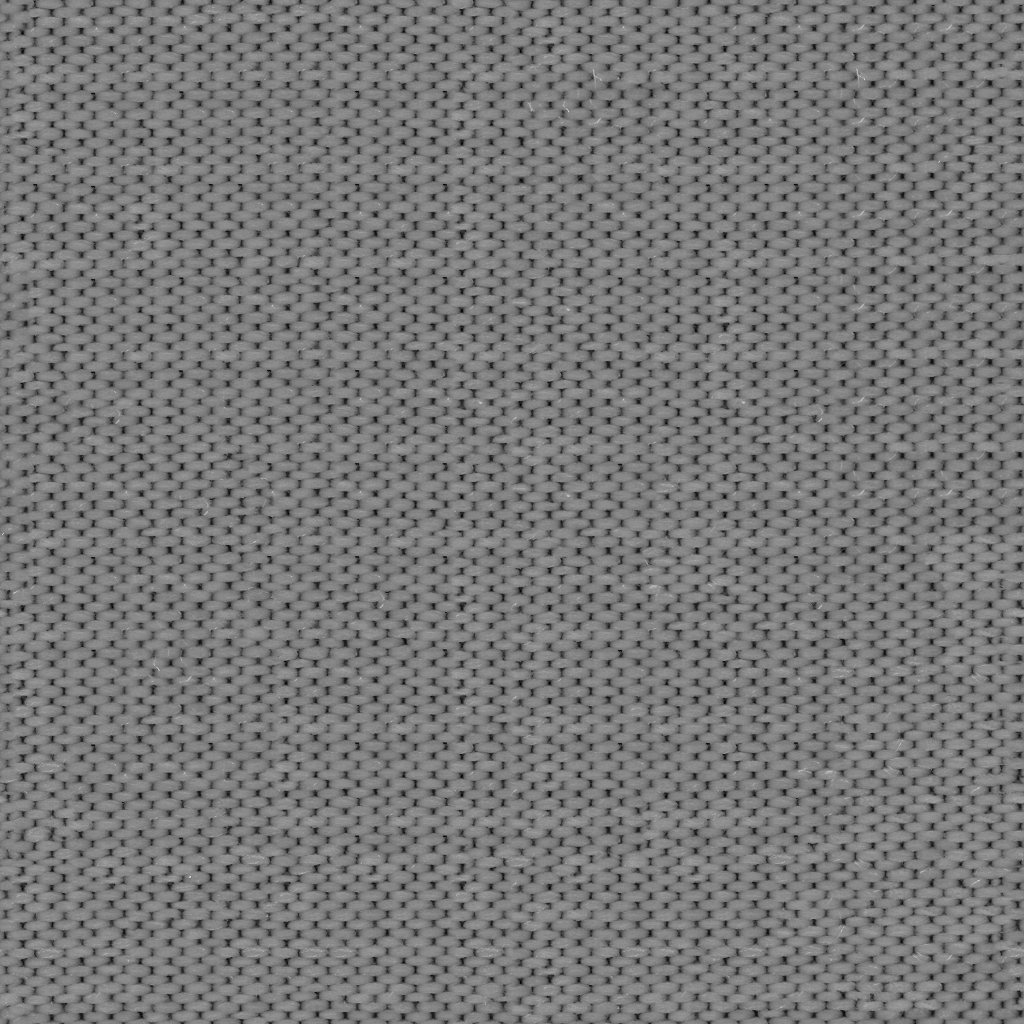}
            \put(795,580){\linethickness{0.2mm}\color{red}\polygon(0,0)(50,0)(50,50)(0,50)}
        \end{overpic} &
        \includegraphics[width=0.15\linewidth]{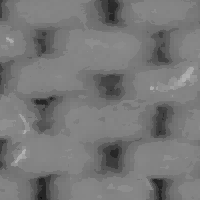} &
        \includegraphics[width=0.15\linewidth]{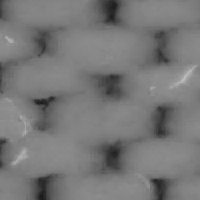} &
        \includegraphics[width=0.15\linewidth]{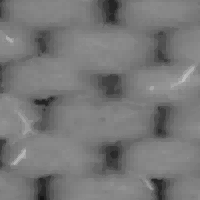} &
        \includegraphics[width=0.15\linewidth]{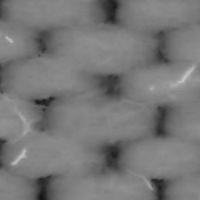} &
        \includegraphics[width=0.15\linewidth]{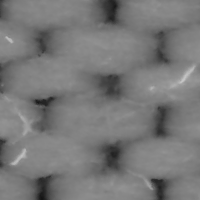} \\


        & PSNR & 34.02 dB & 36.85 dB & 38.53 dB & \textbf{44.20 dB} & 53.06 dB \\
    \end{tabular}
    \caption{Qualitative comparison of the naive approach and NTBC for the diffuse and displacement textures in the \textit{Carpet015} material retrieved from ambientCG~\cite{ambientCG}.
    Diffuse and displacement textures are compressed as BC1 and BC4, respectively.}
    \label{fig:carpet}
\end{figure*}

\paragraph*{Qualitative results.}
Fig.~\ref{fig:carpet} shows compression results of BC1 and BC4 for the diffuse and displacement textures in the \textit{Carpet015} material along with their PSNR values.
For both textures, the naive approach shows visible block artifacts.
NTBC significantly alleviates the artifacts, which results in higher PSNR values, but it also produces a bit blurry images.
We observe the aggressive and conservative approaches show similar results for the diffuse texture, while the conservative approach produces much better results for the displacement texture.
As quantitative results in Fig.~\ref{fig:quantitative} also indicate, the conservative approach can achieve visually smoother results for the single-channel textures.

Fig.~\ref{fig:teaser} also provides other compression results for the \textit{MetalPlates013} material, which has two RGB and four single-channel textures.
Though the PSNR values of NTBC are lower than the reference BC due to a bit blurred results, NTBC shows visually equivalent results, despite using $45\%$ less storage.

Overall, the reference BC with Compressonator shows the best results, but NTBC still provides reasonable results, especially with the conservative approach, with the reduced size of storage.
The standard BC consumes $8$ MB of storage for each texture in the materials, while NTBC with the conservative approach only consumes $26.74$ MB regardless of the number of textures in the materials.

\begin{figure}
    \centering
    \includegraphics[width=0.9\columnwidth]{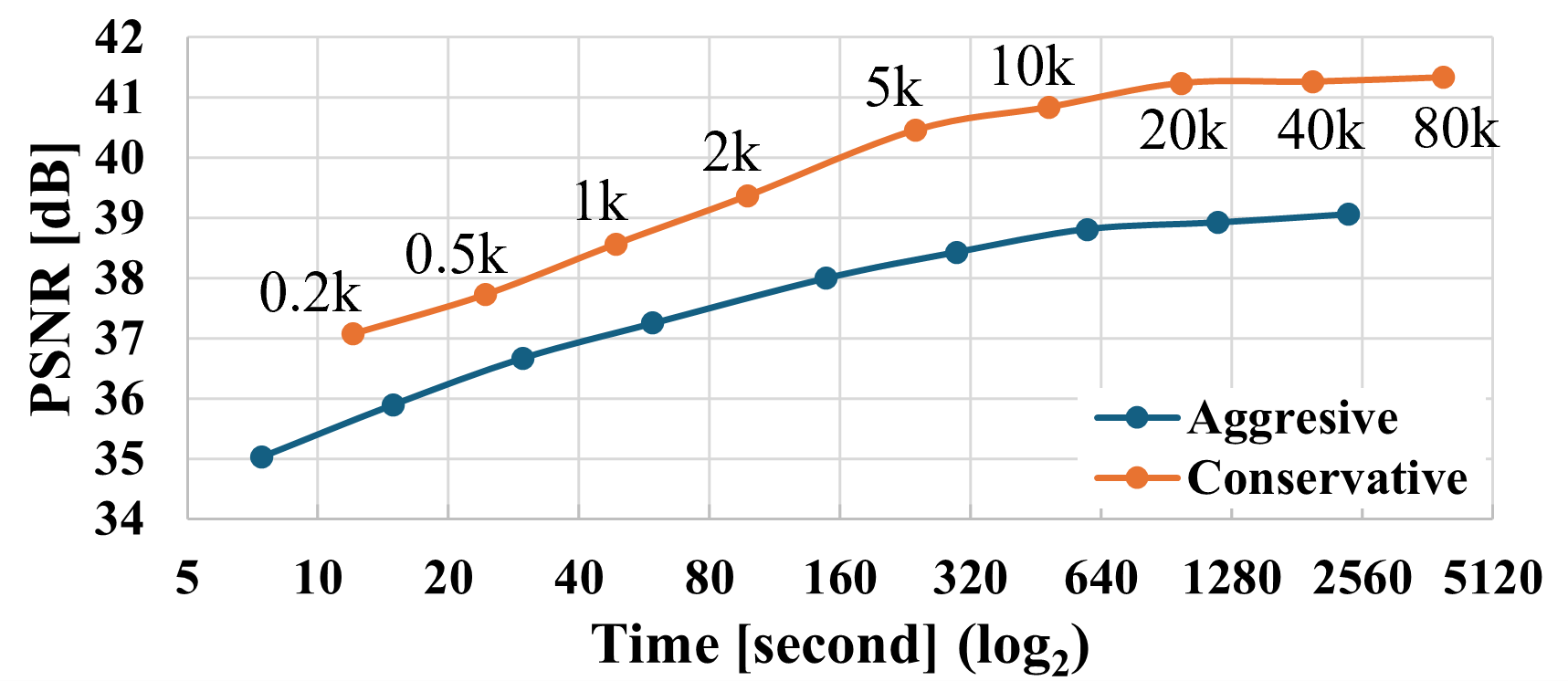}
    \caption{NTBC training times vs. average PSNRs of the conservative and aggressive approaches over six $4$k textures in the \textit{MetalPlates013} material.}
    \label{fig:training}
\end{figure}

\subsection{Performance} \label{sec:performance}
In this section, we demonstrate the training and inference times of our NTBC implementation both for the conservative and aggressive approaches.
We use six $4$k textures in the \textit{MetalPlates013} material from ambientCG~\cite{ambientCG}.

\paragraph*{Training.}
Fig.~\ref{fig:training} shows the training times of NTBC measured for the conservative and aggressive approaches with the PSNR values averaged over all the textures in the material.
The conservative approach requires more training time than the aggressive one because it has to train two distinct models for RGB and single-channel textures.
Almost saturated results can be achieved after $20$k iterations for both approaches, which takes about $10$ minutes and $16$ minutes for the aggressive and conservative approaches, respectively.
The aggressive approach can obtain results with a certain quality in just less than one minute with about $1.5$ dB below the optimal results.

\begin{table}
    \caption{NTBC inference times to predict compressed blocks for \textit{MetalPlates013} material with $2$ BC1 and $4$ BC4 compressed textures.
    Performance is similar for all materials.}
    \label{tab:inferPerf}
    \centering
    \small
    \begin{tabular}{c|c|c|c}
    \toprule
        \multicolumn{2}{c|}{ Conservative } &
        \multicolumn{2}{c}{ Aggressive } \\
        
        BC1 \& BC4 &
        BC1 only &
        BC1 \& BC4 &
        BC1 only \\
    \midrule
        49.84 ms & 25.57 ms & 27.31 ms & 25.96 ms \\
    \bottomrule
    \end{tabular}
\end{table}

\paragraph*{Inference.}
NTBC predicts block-compressed data instead of loading block-compressed textures from the disk, and then texel values are decoded from the compressed data using the existing BC decompression method at runtime.
Therefore, the inference times are simply overhead of NTBC compared to the standard BC.
Tab.~\ref{tab:inferPerf} lists the inference times of our conservative and aggressive approaches to construct compressed blocks of all textures in the material.
The aggressive approach has faster inference than the conservative one because the conservative approach has to run two distinct models for RGB and single-channel textures, which requires four network executions in total.
However, when processing the material containing solely RGB or single-channel textures, the computational workload is reduced to a single model execution.
The results with only two RGB textures are shown in \emph{BC1 only} in Tab.~\ref{tab:inferPerf} where the inference times of the conservative and aggressive approaches are almost equivalent.
Additionally, the \emph{BC1 only} case of the aggressive one shows a similar inference to its \emph{BC1 \& BC4} case, which indicates that NTBC inference remains relatively consistent regardless of the number of textures in the material.
Overall, although NTBC has the computational overhead for the inference of block-compressed data between $27.31$ ms and $49.84$ ms, it is still practical for real-time applications because the inference is executed only once per material instead of loading textures from the disk.

\section{Discussion and Future Work} \label{sec:discussion}
\begin{figure}
    \centering
    \includegraphics[width=0.9\columnwidth]{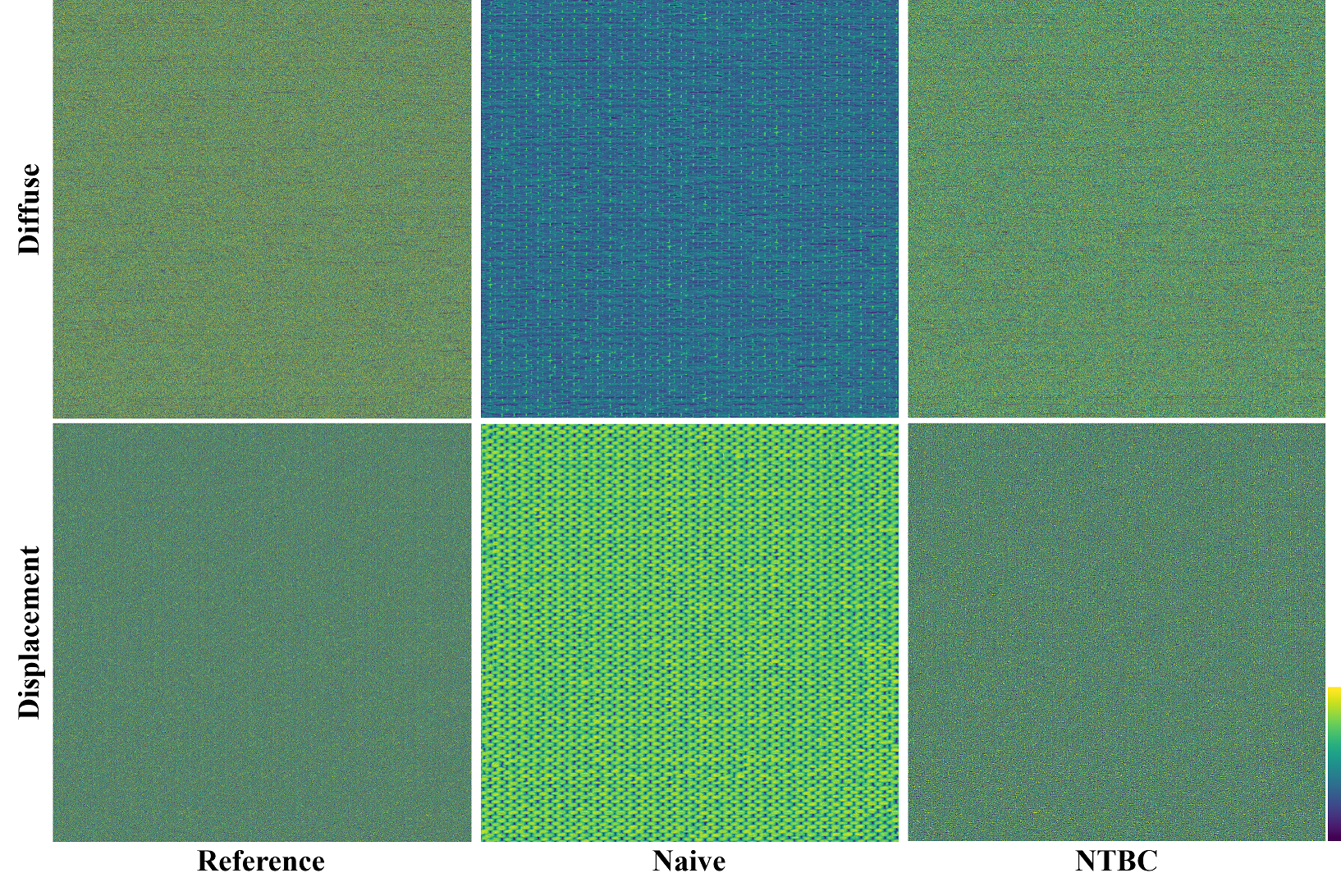}
    \caption{Weight indices of two textures in the \textit{Carpet015} material.
    References are obtained by Compressonator.}
    \label{fig:weightCarpet}
\end{figure}

Our method, NTBC, aims to efficiently compress textures in BC formats with a small storage footprint while maintaining reasonable quality results.
To achieve this goal, NTBC learns the original uncompressed colors of textures instead of their compressed weights.
Fig.~\ref{fig:weightCarpet} visualizes weight indices in the resulting block-compressed data for diffuse and displacement textures in the \textit{Carpet015} material, acquired by three different methods.
The color from black to yellow represents the indices from $0$ to $3$ for diffuse and from $0$ to $7$ for displacement.
The naive approach whose network directly infers weights produces low-frequency results.
On the other hand, NTBC produces high-frequency results that are more equivalent to references with Compressonator, resulting in better-quality BC textures.
Fig.~\ref{fig:teaser} also shows another example of its capability to maintain high-frequency weights for the other material.
In this section, we discuss the limitations of our method and propose future research directions.

\begin{figure}
    \centering
    \begin{subfigure}[b]{0.32\columnwidth}
        \centering
        \includegraphics[width=\textwidth]{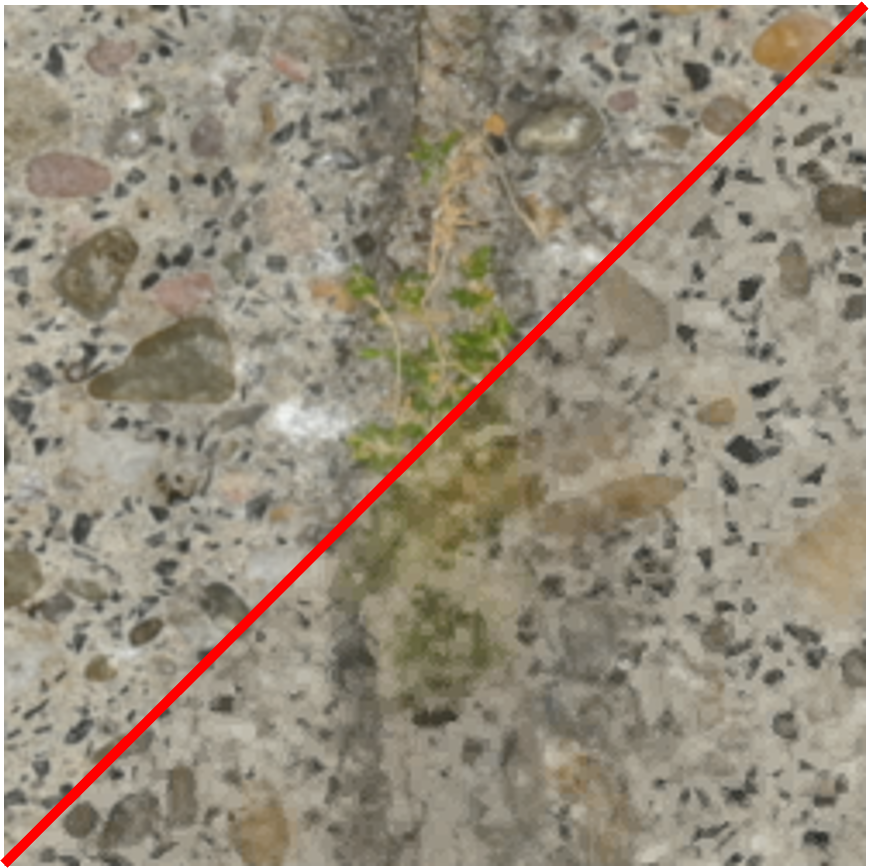}
        \caption{\small{Discoloration}}
        \label{fig:failure_a}
    \end{subfigure}
    \hfill
    \begin{subfigure}[b]{0.32\columnwidth}
        \centering
        \includegraphics[width=\textwidth]{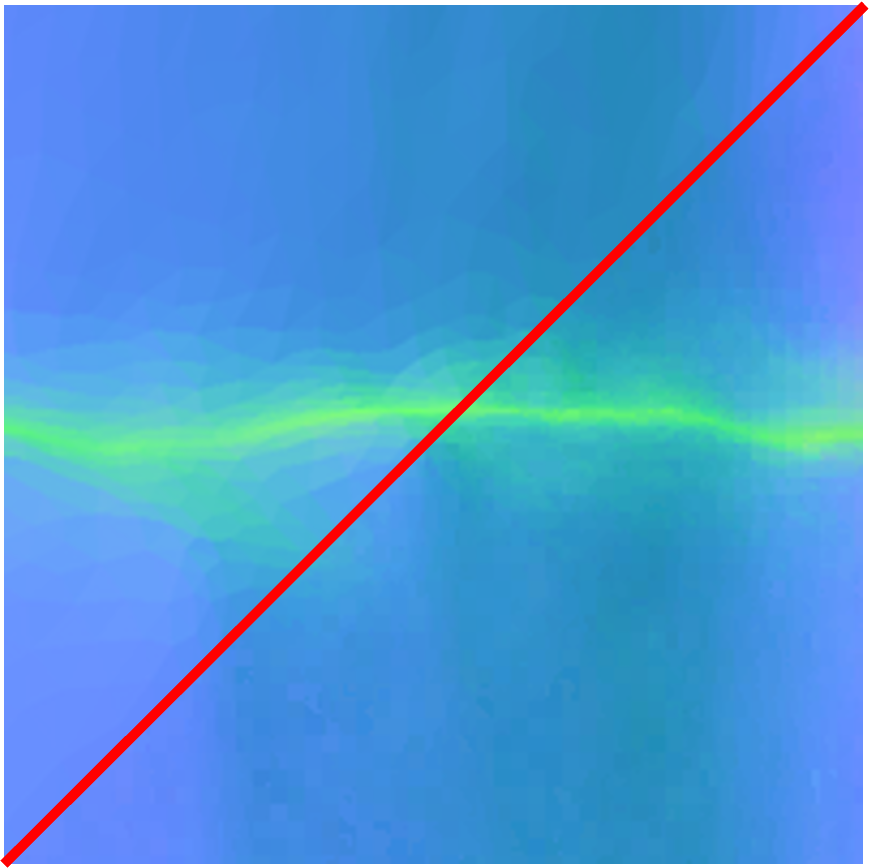}
        \caption{\small{Block artifact}}
        \label{fig:failure_b}
    \end{subfigure}
    \hfill
    \begin{subfigure}[b]{0.32\columnwidth}
        \centering
        \includegraphics[width=\textwidth]{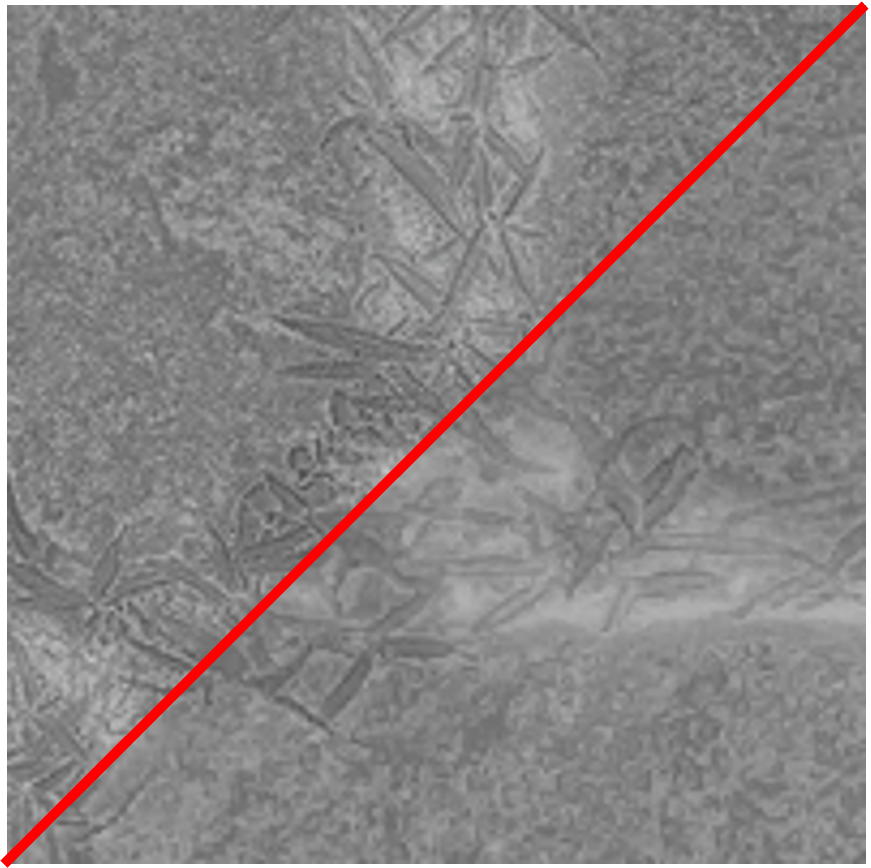}
        \caption{\small{Blurred details}}
        \label{fig:failure_c}
    \end{subfigure}
    \caption{Typical error cases of our method. Top-left: Compressonator, bottom-right: NTBC.}
    \label{fig:failure}
\end{figure}

NTBC sometimes produces errors, such as discoloration, block artifacts, and blurred details, as shown in Fig.~\ref{fig:failure}.
Fig.~\ref{fig:failure_a} shows a discolored texture example which occurs if textures have high-frequency patterns both in contents and colors.
NTBC can achieve good results for high-frequency patterns only in contents, such as normal maps, but if it is also shown in colors, it is challenging to handle both simultaneously.
Fig.~\ref{fig:failure_b} and Fig.~\ref{fig:failure_c} illustrate examples of the loss of details in the compressed textures.
If the texture has smooth gradients, NTBC may produce block artifacts as shown in Fig.~\ref{fig:failure_b}.
And, NTBC may blur the details of the texture if having strong high-frequency patterns as shown in Fig.~\ref{fig:failure_c}.
These errors are caused by the limitation of the NTBC network, which uses lower-resolution grids than the original texture to reduce storage costs.
To address these issues, we can consider other encoding methods, such as texture-focusing encoding~\cite{ntc2023} and local positional encoding~\cite{lpe} using frequency information to handle high-frequency patterns efficiently.
Also, we can use a more sophisticated loss function that considers the perceptual quality of colors in luminance-chrominance space to prevent color degradation.
These are promising and interesting future research directions.

Our experiments described in Sec.~\ref{sec:results} show the storage efficiency of our method for materials with a large number of textures.
The conservative and aggressive approaches of NTBC consume almost fixed sizes of storage, $26.74$ MB and $13.37$ MB, respectively.
On the other hand, the standard BC1 and BC4 formats require $8$ MB for a single $4$k texture.
Therefore, our approach has advantages in terms of storage efficiency if a material contains two or more textures.

In this paper, we focus on BC1 and BC4 which are simple but widely-used formats in real-time applications.
However, these formats only show limited-quality results compared to more complex formats such as BC6H and BC7.
They employ more sophisticated encoding methods, such as block partitioning and flexible bit-rate endpoints using a variety of modes.
We believe that our method can be extended to these formats and leave it as one of interesting future works.

\section{Conclusion} \label{sec:conclusion}
In this paper, we introduced Neural Texture Block Compression (NTBC) which compresses textures in BC formats by learning encoding functions using two MLPs.
It has been designed to be compatible with existing BC formats, especially BC1 and BC4, with the reduced size of storage, and can be used as a drop-in replacement for existing graphics pipelines.
We proposed two different methods to train the encoding functions: one encoding BC1 and BC4 separately (\textit{conservative}) and the other encoding them together in one model (\textit{aggressive}).
We have shown that NTBC can achieve better compression ratios than the standard BC for both approaches while generating reasonable-quality textures.
However, there is a trade-off between quality and compression ratio for the aggressive and conservative approaches, with the aggressive approach achieving better compression ratios at the cost of small quality degradation.
Therefore, we leave it to users to decide which approach, including the standard BC, to use depending on their requirements.
We think this flexibility is one of the strengths of NTBC thanks to its compatibility with the existing BC.

We consider this work as a first step towards learning-based block compression and believe that there is still much potential to improve the quality and compression ratio.
In the future, we plan to explore other input encodings such as local positional encoding~\cite{lpe} and extend NTBC to support more advanced block compression formats such as BC6H and BC7, to achieve more flexibility and better quality.

\subsubsection*{Ackowledgements}
We thank Sylvain Meunier for his inspiring study and insights for this project.
We also thank Joel Jordan at AMD for his feedback and help with the publication.

\bibliographystyle{eg-alpha-doi}

\bibliography{ref}

\newpage

\appendix

\section{Straight-through Estimator for Argmax Operation}
As discussed in Sec.~\ref{sec:ntbc}, we use STE to approximate the gradients of the argmax operation which is not differentiable.
In this section, we discuss the details of this gradient computation considering BC1 for simplicity, and it can be easily extended to BC4 by considering $8$ distances and corresponding weights, described in Sec.~\ref{sec:method}.

We can represent weights for BC1 as Eq.~\ref{eq:quantized_weight}:
\begin{equation} \label{eq:quantized_weight}
    w_n = \frac{n}{3},
\end{equation}
where $n$ is a $2$-bit index per pixel ($0 \leq n \leq 3$).
Distances $d_n$ represent the similarity between four colors on the palette and an uncompressed color.
Thus, the probability $\sigma(d_n)$ in Eq.~\ref{eq:softmax} acts as the likelihood of each corresponding weight $w_n$ in Eq.~\ref{eq:quantized_weight}.
Then, we can achieve the optimal weight $w_n$ by taking one distance with the maximum likelihood using the argmax operation.
Instead, in the backward pass, we consider that the resulting weight was computed as the probability-weighted sum which is equivalent to the expectation of the weights:
\begin{align}
    \hat{w_n} &= \sum_{i \in n} w_i \sigma(d_i), \\
              &= \frac{1}{3} \sigma(d_1) + \frac{2}{3} \sigma(d_2) + \sigma(d_3),
    \label{eq:expected_weight}
\end{align}
with Eq.~\ref{eq:softmax} and Eq.~\ref{eq:quantized_weight}.

The derivative of the softmax operation $\sigma(d_n)$ can be computed as Eq.~\ref{eq:dsoftmax}:
\begin{equation}
    \label{eq:dsoftmax}
    \frac{\partial \sigma(d_i)}{\partial d_j} =
        \begin{cases}
            \frac{1}{T} \cdot \sigma(d_i) \cdot (1 - \sigma(d_j)) & (i = j) \\
            -\frac{1}{T} \cdot \sigma(d_i) \cdot \sigma(d_j) & ( i \neq j )
        \end{cases}
\end{equation}
Then, considering Eq.~\ref{eq:expected_weight} and Eq.~\ref{eq:dsoftmax}, the derivatives of the expected weight $\hat{w_n}$ with respect to each distance are computed as following equations:
\begin{align}
    \frac{\partial \hat{w_n}}{\partial d_0} = {}& \sum_{i \in n} \frac{\partial \hat{w_n}}{\partial \sigma(d_i)} \cdot \frac{\partial \sigma(d_i)}{\partial d_0} \\
    \begin{split}
        = {}& - \frac{1}{3} \cdot \frac{1}{T} \cdot \sigma(d_1) \cdot \sigma(d_0) \\
          {}& - \frac{2}{3} \cdot \frac{1}{T} \cdot \sigma(d_2) \cdot \sigma(d_0) - \frac{1}{T} \cdot \sigma(d_3) \cdot \sigma(d_0)
    \end{split} \\
    = {}& \frac{1}{T} \cdot \sigma(d_0) \cdot (0 - \hat{w_n}),
    \label{eq:dsoftargmin0}
\end{align}
\begin{align}
    \frac{\partial \hat{w_n}}{\partial d_1} = {}& \sum_{i \in n} \frac{\partial \hat{w_n}}{\partial \sigma(d_i)} \cdot \frac{\partial \sigma(d_i)}{\partial d_1} \\
    \begin{split}
        = {}& \frac{1}{3} \cdot \frac{1}{T} \cdot \sigma(d_1) \cdot (1 - \sigma(d_1)) \\
          {}& - \frac{2}{3} \cdot \frac{1}{T} \cdot \sigma(d_2) \cdot \sigma(d_1) - \frac{1}{T} \cdot \sigma(d_3) \cdot \sigma(d_1)
    \end{split} \\
    = {}& \frac{1}{T} \cdot \sigma(d_1)\cdot (1 - \hat{w_n}),
    \label{eq:dsoftargmin1}
\end{align}
\begin{align}
    \frac{\partial \hat{w_n}}{\partial d_2} = {}& \sum_{i \in n} \frac{\partial \hat{w_n}}{\partial \sigma(d_i)} \cdot \frac{\partial \sigma(d_i)}{\partial d_2} \\
    \begin{split}
        = {}& - \frac{1}{3} \cdot \frac{1}{T} \cdot \sigma(d_1) \cdot \sigma(d_2) \\
          {}& + \frac{2}{3} \cdot \frac{1}{T} \cdot \sigma(d_2) \cdot (1 - \sigma(d_2)) - \frac{1}{T} \cdot \sigma(d_3) \cdot \sigma(d_2)
    \end{split} \\
    = {}& \frac{1}{T} \cdot \sigma(d_2) \cdot (2 - \hat{w_n}),
    \label{eq:dsoftargmin2}
\end{align}
\begin{align}
    \frac{\partial \hat{w_n}}{\partial d_3} = {}& \sum_{i \in n} \frac{\partial \hat{w_n}}{\partial \sigma(d_i)} \cdot \frac{\partial \sigma(d_i)}{\partial d_3} \\
    \begin{split}
        = {}& - \frac{1}{3} \cdot \frac{1}{T} \cdot \sigma(d_1) \cdot \sigma(d_3) - \frac{2}{3} \cdot \frac{1}{T} \cdot \sigma(d_2) \cdot \sigma(d_3) \\
          {}& + \frac{1}{T} \cdot \sigma(d_3) \cdot (1 - \sigma(d_3))
    \end{split} \\
    = {}& \frac{1}{T} \cdot \sigma(d_3) \cdot (3 - \hat{w_n}).
    \label{eq:dsoftargmin3}
\end{align}
Therefore, with Eq.~\ref{eq:dsoftargmin0}, Eq.~\ref{eq:dsoftargmin1}, Eq.~\ref{eq:dsoftargmin2}, and Eq.~\ref{eq:dsoftargmin3}, we can summarize the derivatives of the expected weight as
\begin{equation} \label{eq:devExtW}
    \frac{\partial \hat{w_n}}{\partial d_n} = \frac{1}{T} \cdot \sigma(d_n) \cdot (n - \hat{w_n}).
\end{equation}
Using the gradients in Eq.~\ref{eq:devExtW} in the backward pass, the networks can be optimized with the loss $\mathcal{L}_{cd}$.

\section{Additional Results}
\begin{table}[t]
    \centering
    \scriptsize
    \caption{Qualitative comparison with PSNR for materials from ambientCG~\cite{ambientCG}.
    \textit{Ref. BC} shows the reference results with Compressonator.
    The bold numbers indicate the best results for each texture among NN-based approaches.}
    \begin{tabular}{c c||c|c c|c c}
    \toprule
    & & & \multicolumn{2}{c|}{ Aggressive } & \multicolumn{2}{c}{ Conservative } \\
    & & Ref. BC & Naive & NTBC & Naive & NTBC \\
    \midrule
    & Size & 40 / 48 MB & \multicolumn{2}{c|}{ 13.37 MB } & \multicolumn{2}{c}{ 26.74 MB } \\
    \midrule
    \multirow{5}{1.3mm}{\raisebox{0.08\linewidth}{\rotatebox[origin = c]{90}{Bricks090}}}
    & Diffuse & 39.43 & 31.47 & \textbf{34.92} & 31.59 & 34.90 \\
    & Normal & 35.48 & 32.22 & 32.48 & 32.50 & \textbf{32.86} \\[1pt] \cline{2-7} &&&&&& \\[-4pt]
    & Displacement & 52.55 & 42.96 & 44.74 & 47.53 & \textbf{49.42} \\
    & Roughness & 47.42 & 33.02 & 34.80 & 35.84 & \textbf{36.66} \\
    & AO & 51.95 & 35.03 & 35.65 & 38.97 & \textbf{39.85} \\
    \midrule
    \multirow{5}{1.3mm}{\raisebox{0.08\linewidth}{\rotatebox[origin = c]{90}{Carpet015}}}
    & Diffuse & 37.44 & 28.21 & 30.90 & 27.81 & \textbf{30.98} \\
    & Normal & 29.18 & 26.40 & 26.59 & 26.67 & \textbf{26.78} \\[1pt] \cline{2-7} &&&&&& \\[-4pt]
    & Displacement & 53.06 & 34.02 & 36.85 & 38.53 & \textbf{44.20} \\
    & Roughness & 49.45 & 35.17 & 35.86 & 37.64 & \textbf{38.89} \\
    & AO & 47.88 & 31.04 & 31.98 & 34.16 & \textbf{39.04} \\
    \midrule
    \multirow{6}{1.3mm}{\raisebox{0.08\linewidth}{\rotatebox[origin = c]{90}{MetalPlates013}}}
    & Diffuse & 42.53 & 35.46 & 37.76 & 36.10 & \textbf{38.64} \\
    & Normal & 38.34 & 34.02 & 33.92 & 35.26 & \textbf{35.38} \\[1pt] \cline{2-7} &&&&&& \\[-4pt]
    & Displacement & 56.09 & 43.68 & 44.44 & 45.80 & \textbf{48.08} \\
    & Roughness & 49.39 & 33.40 & 36.20 & 37.16 & \textbf{38.87} \\
    & AO & 67.84 & 45.75 & 46.22 & 47.77 & \textbf{50.61} \\
    & Metalness & 46.73 & 31.99 & 33.25 & 33.66 & \textbf{34.26} \\
    \midrule
    \raisebox{0.005\linewidth}{\multirow{5}{1.3mm}{\raisebox{0.08\linewidth}{\rotatebox[origin = c]{90}{PavingStones070}}}}
    & Diffuse & 31.20 & 23.59 & 24.90 & 23.95 & \textbf{25.12} \\
    & Normal & 27.44 & 24.51 & 24.10 & \textbf{24.65} & 24.48 \\[1pt] \cline{2-7} &&&&&& \\[-4pt]
    & Displacement & 53.55 & 35.21 & 37.45 & 43.18 & \textbf{44.01} \\
    & Roughness & 45.77 & 28.01 & 28.51 & \textbf{32.91} & 32.76 \\
    & AO & 46.44 & 30.90 & 31.16 & 34.94 & \textbf{36.91} \\
    \midrule
    \multirow{6}{1.3mm}{\raisebox{0.08\linewidth}{\rotatebox[origin = c]{90}{Rails001}}}
    & Diffuse & 35.90 & 27.16 & 29.85 & 27.20 & \textbf{30.16} \\
    & Normal & 27.09 & 24.89 & 24.54 & \textbf{24.90} & 24.72 \\[1pt] \cline{2-7} &&&&&& \\[-4pt]
    & Displacement & 53.11 & 41.41 & 41.57 & 46.53 & \textbf{47.97} \\
    & Roughness & 52.69 & 33.71 & 33.81 & 40.76 & \textbf{43.71} \\
    & AO & 49.81 & 32.78 & 32.60 & 39.08 & \textbf{40.99} \\
    & Metalness & 68.63 & 44.55 & 47.20 & 47.33 & \textbf{53.34} \\
    \midrule
    \multirow{5}{1.3mm}{\raisebox{0.08\linewidth}{\rotatebox[origin = c]{90}{Wood063}}}
    & Diffuse & 33.79 & 26.02 & 27.90 & 26.12 & \textbf{28.22} \\
    & Normal & 26.81 & 24.30 & 24.15 & 24.58 & \textbf{24.65} \\[1pt] \cline{2-7} &&&&&& \\[-4pt]
    & Displacement & 48.40 & 30.35 & 34.29 & 35.04 & \textbf{41.26} \\
    & Roughness & 44.73 & 28.58 & 29.60 & 31.81 & \textbf{37.65} \\
    & AO & 46.12 & 31.26 & 31.22 & 34.92 & \textbf{35.13} \\
    \bottomrule
    \end{tabular}
    \label{tab:ambientCG}
\end{table}

\begin{table}[t]
    \centering
    \scriptsize
    \caption{Qualitative comparison with PSNR for materials from Poly Haven~\cite{polyhaven}.
    \textit{Ref. BC} shows the reference results with Compressonator.
    The bold numbers indicate the best results for each texture among NN-based approaches.}
    \begin{tabular}{c c||c|c c|c c}
    \toprule
    & & & \multicolumn{2}{c|}{ Aggressive } & \multicolumn{2}{c}{ Conservative } \\
    & & Ref. BC & Naive & NTBC & Naive & NTBC \\
    \midrule
    & Size & 40 MB & \multicolumn{2}{c|}{ 13.37 MB } & \multicolumn{2}{c}{ 26.74 MB } \\
    \midrule
    \multirow{5}{1.3mm}{\raisebox{0.08\linewidth}{\rotatebox[origin = c]{90}{aerial\_rocks\_02}}}
    & Diffuse & 36.82 & 28.31 & 30.41 & 28.46 & \textbf{30.74} \\
    & Normal & 29.01 & 26.25 & 26.12 & \textbf{26.41} & \textbf{26.4}1 \\[1pt] \cline{2-7} &&&&&& \\[-4pt]
    & Displacement & 56.62 & 41.18 & 42.33 & 47.55 & \textbf{48.77} \\
    & Roughness & 49.20 & 34.16 & 36.93 & 36.63 & \textbf{39.21} \\
    & AO & 47.32 & 30.03 & 31.11 & 34.82 & \textbf{36.48} \\
    \midrule
    \multirow{5}{1.3mm}{\raisebox{0.08\linewidth}{\rotatebox[origin = c]{90}{forest\_sand\_01}}}
    & Diffuse & 31.39 & 23.71 & \textbf{26.00} & 24.04 & 25.95 \\
    & Normal & 31.82 & \textbf{27.53} & 27.20 & 27.30 & 27.20 \\
    & ARM & 38.29 & 29.96 & 30.92 & 29.79 & \textbf{31.01} \\[1pt] \cline{2-7} &&&&&& \\[-4pt]
    & Displacement & 55.79 & 41.39 & 40.79 & 45.98 & \textbf{48.45} \\
    & Specular & 45.21 & 31.30 & 31.37 & \textbf{35.31} & 35.27 \\
    \midrule
    \raisebox{0.005\linewidth}{\multirow{5}{1.3mm}{\raisebox{0.08\linewidth}{\rotatebox[origin = c]{90}{red\_dirt\_mud\_01}}}}
    & Diffuse & 35.89 & 27.65 & 31.62 & 27.82 & \textbf{31.91} \\
    & Normal & 35.20 & 30.36 & 30.20 & \textbf{30.40} & 30.03 \\
    & ARM & 34.19 & 28.40 & 31.13 & 28.67 & \textbf{31.35} \\[1pt] \cline{2-7} &&&&&& \\[-4pt]
    & Displacement & 53.56 & 36.59 & 39.28 & 44.03 & \textbf{48.99} \\
    & Specular & 51.41 & 39.57 & 41.04 & 42.73 & \textbf{44.45} \\
    \midrule
    \multirow{5}{1.3mm}{\raisebox{0.08\linewidth}{\rotatebox[origin = c]{90}{roof\_09}}}
    & Diffuse & 40.29 & 32.39 & 37.34 & 32.51 & \textbf{37.44} \\
    & Normal & 45.05 & 37.53 & 38.61 & 37.73 & \textbf{38.65} \\
    & ARM & 39.50 & 32.87 & 37.40 & 32.97 & \textbf{37.78} \\[1pt] \cline{2-7} &&&&&& \\[-4pt]
    & Displacement & 55.52 & 41.53 & 45.29 & 44.20 & \textbf{50.32} \\
    & Specular & 52.02 & 41.81 & 43.63 & 44.22 & \textbf{46.40} \\
    \bottomrule
    \end{tabular}
    \label{tab:PolyHaven}
\end{table}

Tab.~\ref{tab:ambientCG} and Tab.~\ref{tab:PolyHaven} list the PSNR values for all the materials in our dataset retrieved from ambientCG~\cite{ambientCG} and Poly Haven~\cite{polyhaven}, respectively.
PSNR values are calculated between the original uncompressed textures and the decompressed textures from block-compressed data achieved with the naive approach, NTBC, and Compressonator~\cite{Compressonator}.
Compressonator uses two refine steps for BC1 compression, and the naive approach and NTBC are trained for $20$k steps.
We compress diffuse, normal, and ARM textures as BC1 and other textures as BC4.

The results show that conservative NTBC achieves the best PSNR values for most textures among all NN-based methods, and NTBC outperforms the naive approach in most cases.
Comparing the aggressive and conservative approaches for each method, they have similar results for BC1-compressed textures while the conservative approach shows a significant improvement for BC4-compressed textures.
BC1-compressed textures have more spatial information than BC4-compressed textures thanks to multiple channels and show relatively lower PSNR values even compressed with Compressonator, which results in a smaller gap between the aggressive and conservative approaches.
Also, focusing on normal textures, even the naive approach can obtain relatively high PSNR, leading to similar results for all methods.
Normal textures have similar endpoints across the texture, which makes weights relatively lower-frequency and easier to compress.
So, we recommend users select the proper approach depending on the use cases to achieve the best results.
Higher-quality formats like BC7 and other more texture-specific input encodings could improve the quality of these textures further, but we leave them for our future work.

\begin{figure*}
    \centering
    \setlength{\tabcolsep}{0\linewidth}
    \renewcommand{\arraystretch}{1.0}
    \begin{tabular}{c@{\hspace{0.004\linewidth}} c@{\hspace{0.002\linewidth}} |@{\hspace{0.002\linewidth}} c@{\hspace{0.002\linewidth}} |@{\hspace{0.002\linewidth}} c@{\hspace{0.002\linewidth}} c@{\hspace{0.002\linewidth}} |@{\hspace{0.002\linewidth}} c@{\hspace{0.002\linewidth}} c}
        & & & \multicolumn{2}{c@{\hspace{0.002\linewidth}}|@{\hspace{0.002\linewidth}}}{ Aggressive } & \multicolumn{2}{c}{ Conservative } \\

        & & Ref. BC & Naive & NTBC & Naive & NTBC \\


        & Size & 40 MB & \multicolumn{2}{c@{\hspace{0.002\linewidth}}|@{\hspace{0.002\linewidth}}}{ 13.37 MB } & \multicolumn{2}{c}{ 26.74 MB } \\

        \midrule

        \raisebox{0.08\linewidth}{\rotatebox[origin = c]{90}{Diffuse}} &
        \includegraphics[width=0.15\linewidth]{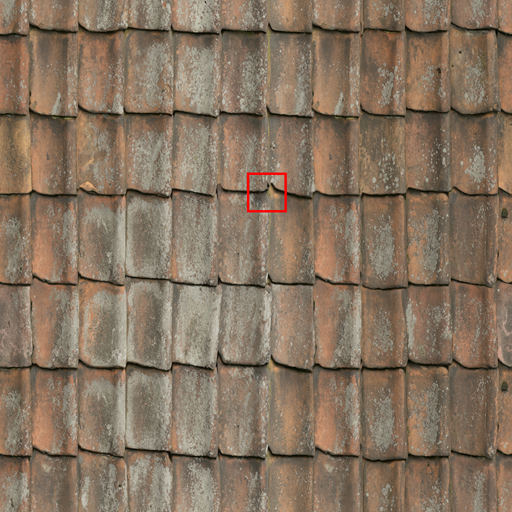} &
        \includegraphics[width=0.15\linewidth]{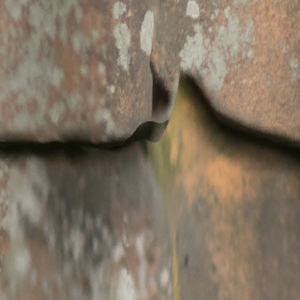} &
        \includegraphics[width=0.15\linewidth]{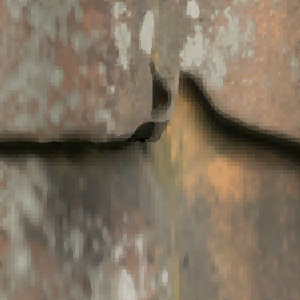} &
        \includegraphics[width=0.15\linewidth]{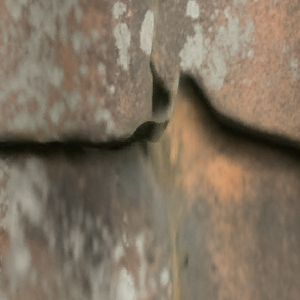} &
        \includegraphics[width=0.15\linewidth]{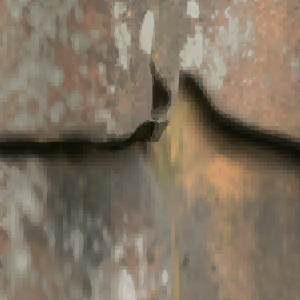} &
        \includegraphics[width=0.15\linewidth]{figures/roofComp/NTBC_CS_diffuse_crop.png} \\

        \midrule

        \raisebox{0.08\linewidth}{\rotatebox[origin = c]{90}{Normal}} &
        \includegraphics[width=0.15\linewidth]{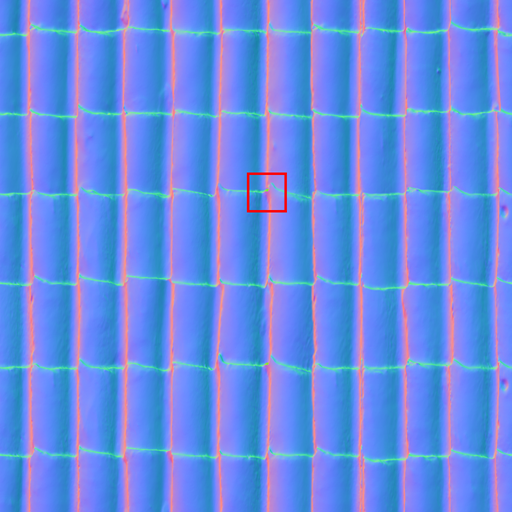} &
        \includegraphics[width=0.15\linewidth]{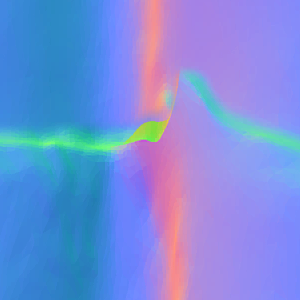} &
        \includegraphics[width=0.15\linewidth]{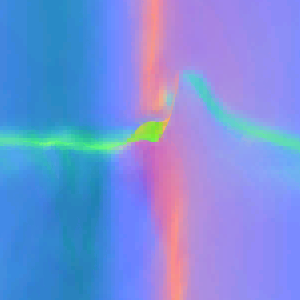} &
        \includegraphics[width=0.15\linewidth]{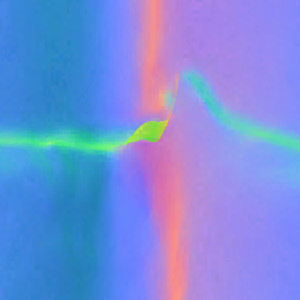} &
        \includegraphics[width=0.15\linewidth]{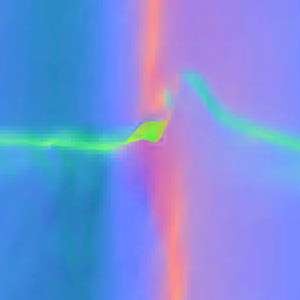} &
        \includegraphics[width=0.15\linewidth]{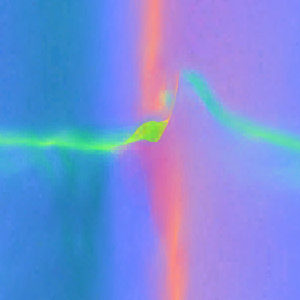} \\

        \midrule

        \raisebox{0.08\linewidth}{\rotatebox[origin = c]{90}{ARM}} &
        \includegraphics[width=0.15\linewidth]{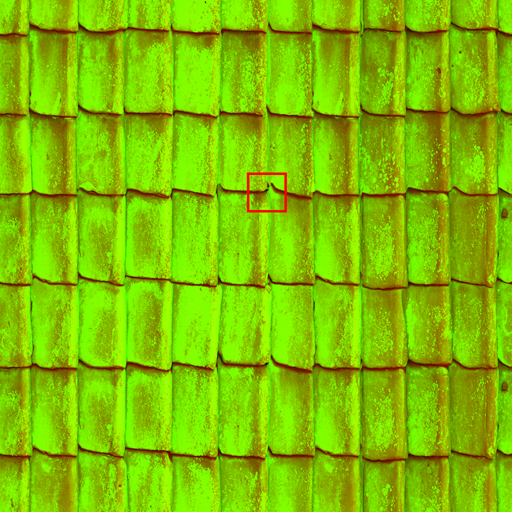} &
        \includegraphics[width=0.15\linewidth]{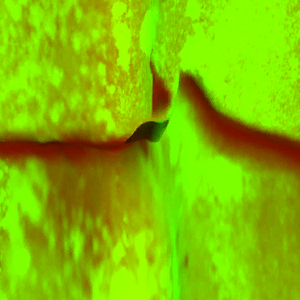} &
        \includegraphics[width=0.15\linewidth]{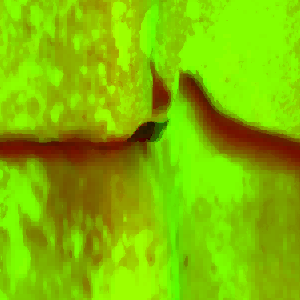} &
        \includegraphics[width=0.15\linewidth]{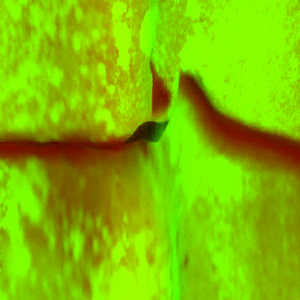} &
        \includegraphics[width=0.15\linewidth]{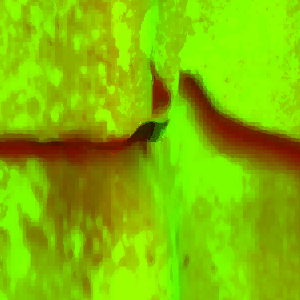} &
        \includegraphics[width=0.15\linewidth]{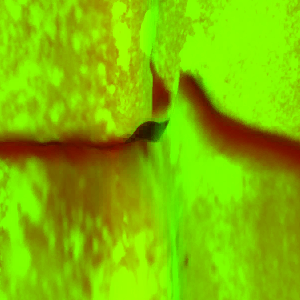} \\

        \midrule

        \raisebox{0.08\linewidth}{\rotatebox[origin = c]{90}{Displacement}} &
        \includegraphics[width=0.15\linewidth]{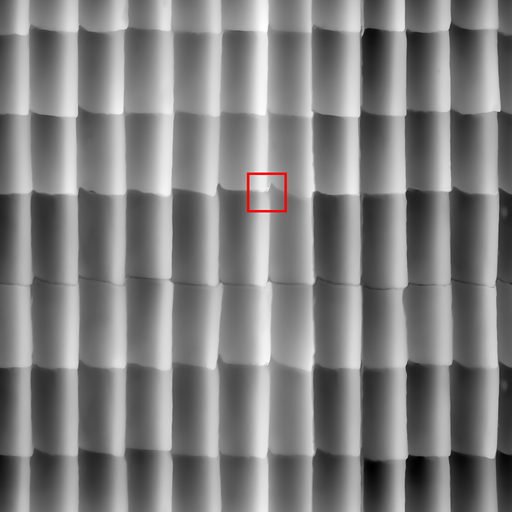} &
        \includegraphics[width=0.15\linewidth]{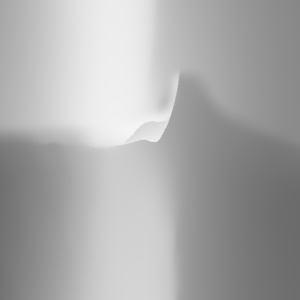} &
        \includegraphics[width=0.15\linewidth]{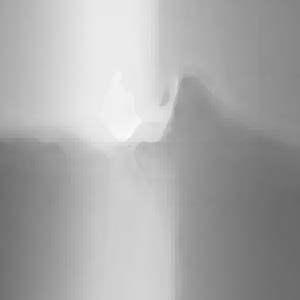} &
        \includegraphics[width=0.15\linewidth]{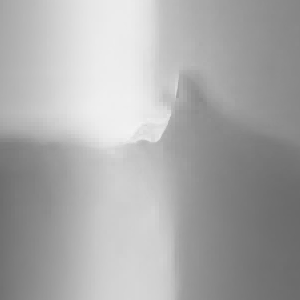} &
        \includegraphics[width=0.15\linewidth]{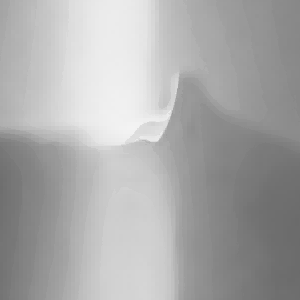} &
        \includegraphics[width=0.15\linewidth]{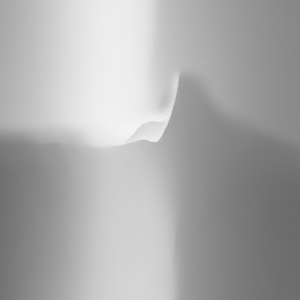} \\

        \midrule

        \raisebox{0.08\linewidth}{\rotatebox[origin = c]{90}{Specular}} &
        \includegraphics[width=0.15\linewidth]{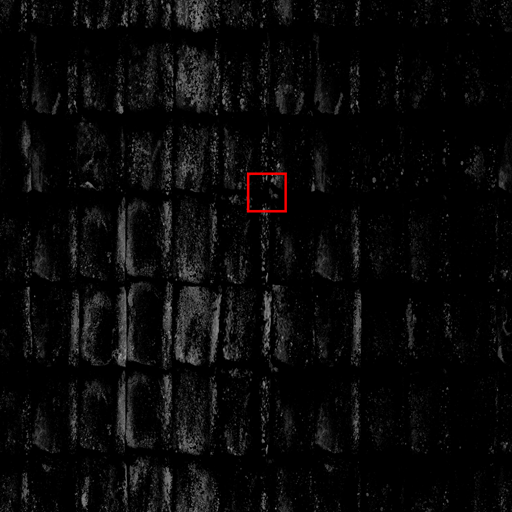} &
        \includegraphics[width=0.15\linewidth]{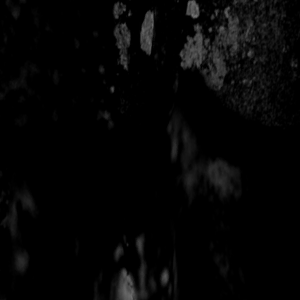} &
        \includegraphics[width=0.15\linewidth]{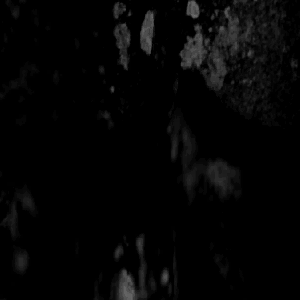} &
        \includegraphics[width=0.15\linewidth]{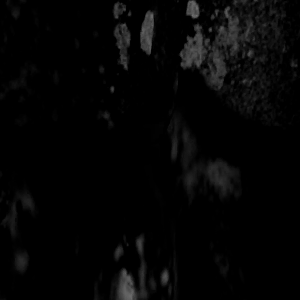} &
        \includegraphics[width=0.15\linewidth]{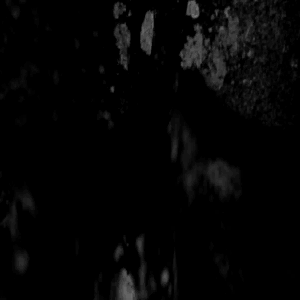} &
        \includegraphics[width=0.15\linewidth]{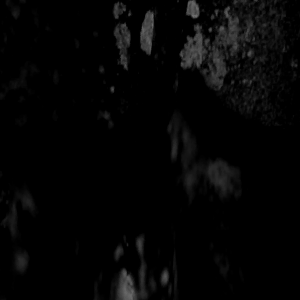} \\
    \end{tabular}
    \caption{Comparison of different methods for the \textit{roof\_09} material, retrieved from Poly Haven~\cite{polyhaven}.
    All methods show relatively higher PSNR values for this material.}
    \label{fig:roof}
\end{figure*}

\begin{figure*}
    \centering
    \setlength{\tabcolsep}{0\linewidth}
    \renewcommand{\arraystretch}{1.0}
    \begin{tabular}{c@{\hspace{0.004\linewidth}} c@{\hspace{0.002\linewidth}} |@{\hspace{0.002\linewidth}} c@{\hspace{0.002\linewidth}} |@{\hspace{0.002\linewidth}} c@{\hspace{0.002\linewidth}} c@{\hspace{0.002\linewidth}} |@{\hspace{0.002\linewidth}} c@{\hspace{0.002\linewidth}} c}
        & & & \multicolumn{2}{c@{\hspace{0.002\linewidth}}|@{\hspace{0.002\linewidth}}}{ Aggressive } & \multicolumn{2}{c}{ Conservative } \\

        & & Ref. BC & Naive & NTBC & Naive & NTBC \\


        & Size & 40 MB & \multicolumn{2}{c@{\hspace{0.002\linewidth}}|@{\hspace{0.002\linewidth}}}{ 13.37 MB } & \multicolumn{2}{c}{ 26.74 MB } \\

        \midrule

        \raisebox{0.08\linewidth}{\rotatebox[origin = c]{90}{Diffuse}} &
        \includegraphics[width=0.15\linewidth]{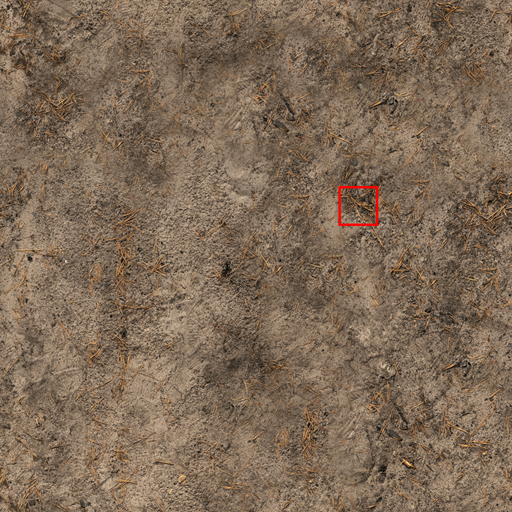} &
        \includegraphics[width=0.15\linewidth]{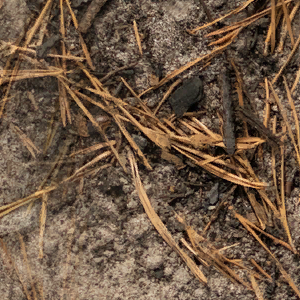} &
        \includegraphics[width=0.15\linewidth]{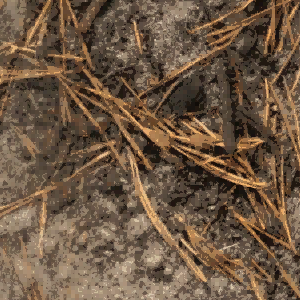} &
        \includegraphics[width=0.15\linewidth]{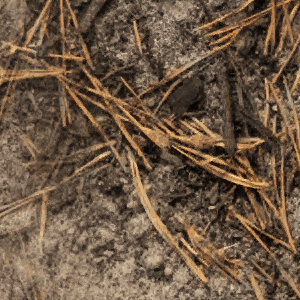} &
        \includegraphics[width=0.15\linewidth]{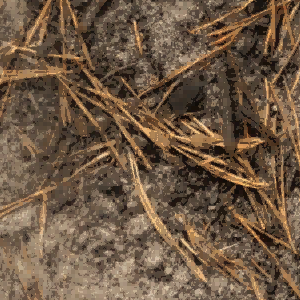} &
        \includegraphics[width=0.15\linewidth]{figures/forestComp/NTBC_CS_diffuse_crop.png} \\

        \midrule

        \raisebox{0.08\linewidth}{\rotatebox[origin = c]{90}{Normal}} &
        \includegraphics[width=0.15\linewidth]{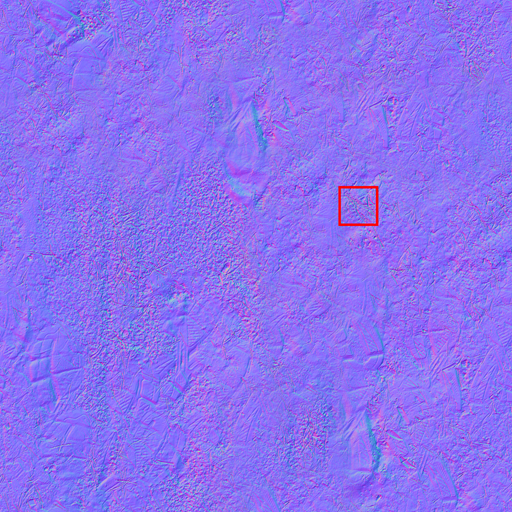} &
        \includegraphics[width=0.15\linewidth]{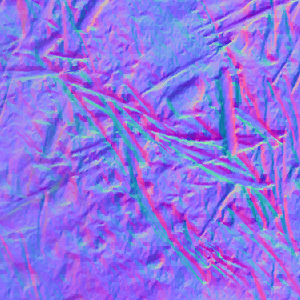} &
        \includegraphics[width=0.15\linewidth]{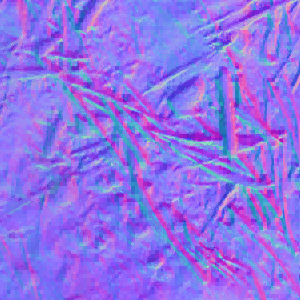} &
        \includegraphics[width=0.15\linewidth]{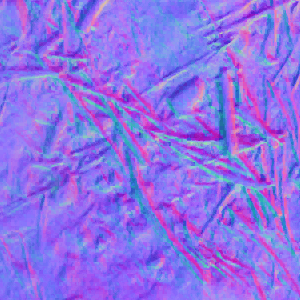} &
        \includegraphics[width=0.15\linewidth]{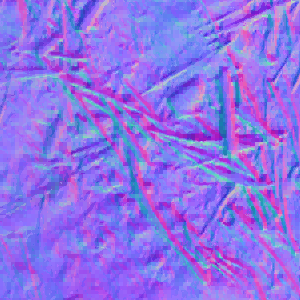} &
        \includegraphics[width=0.15\linewidth]{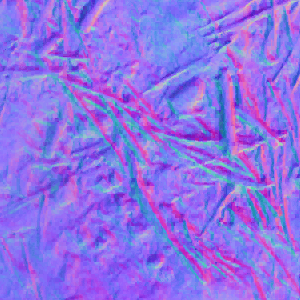} \\

        \midrule

        \raisebox{0.08\linewidth}{\rotatebox[origin = c]{90}{ARM}} &
        \includegraphics[width=0.15\linewidth]{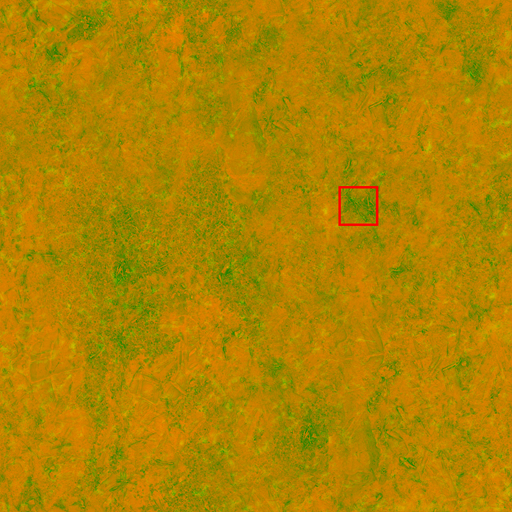} &
        \includegraphics[width=0.15\linewidth]{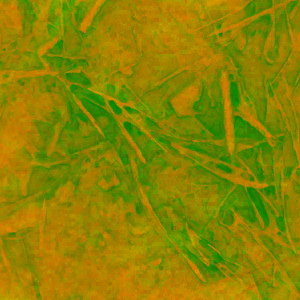} &
        \includegraphics[width=0.15\linewidth]{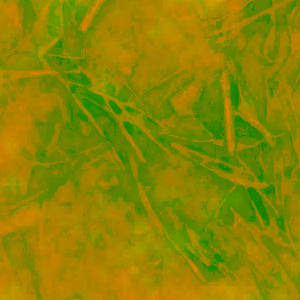} &
        \includegraphics[width=0.15\linewidth]{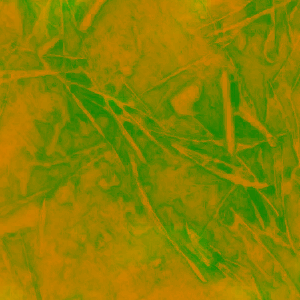} &
        \includegraphics[width=0.15\linewidth]{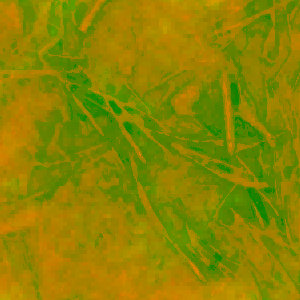} &
        \includegraphics[width=0.15\linewidth]{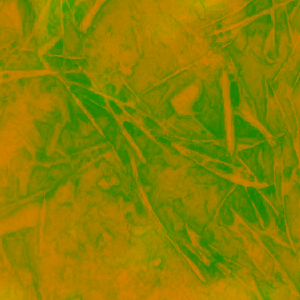} \\

        \midrule

        \raisebox{0.08\linewidth}{\rotatebox[origin = c]{90}{Displacement}} &
        \includegraphics[width=0.15\linewidth]{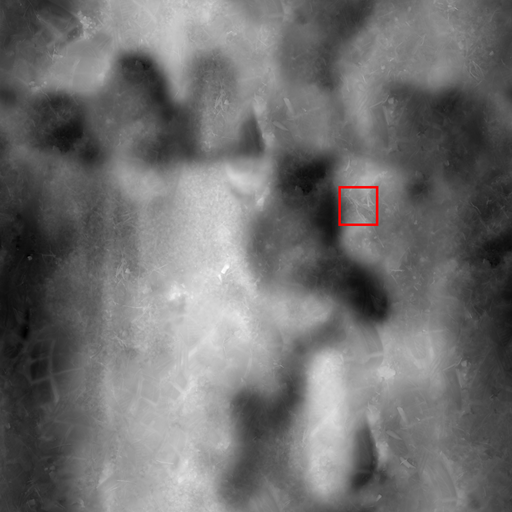} &
        \includegraphics[width=0.15\linewidth]{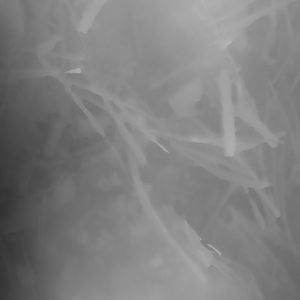} &
        \includegraphics[width=0.15\linewidth]{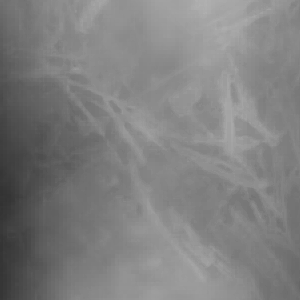} &
        \includegraphics[width=0.15\linewidth]{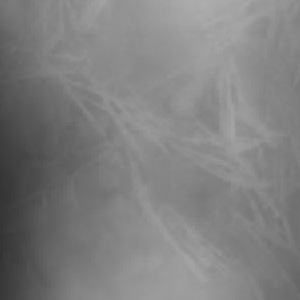} &
        \includegraphics[width=0.15\linewidth]{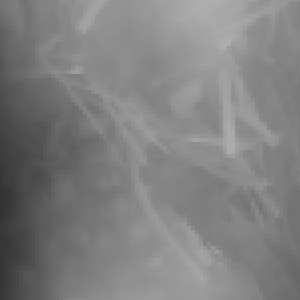} &
        \includegraphics[width=0.15\linewidth]{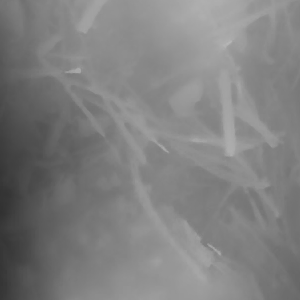} \\

        \midrule

        \raisebox{0.08\linewidth}{\rotatebox[origin = c]{90}{Specular}} &
        \includegraphics[width=0.15\linewidth]{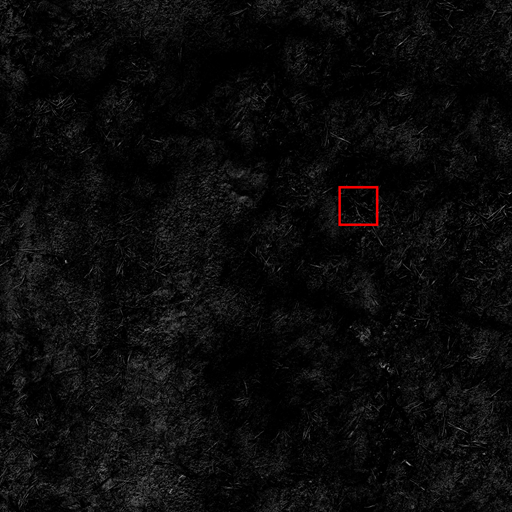} &
        \includegraphics[width=0.15\linewidth]{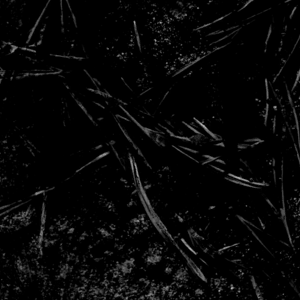} &
        \includegraphics[width=0.15\linewidth]{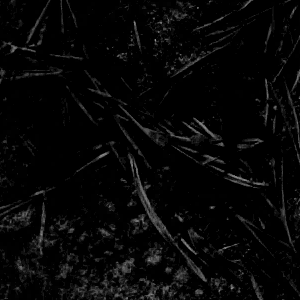} &
        \includegraphics[width=0.15\linewidth]{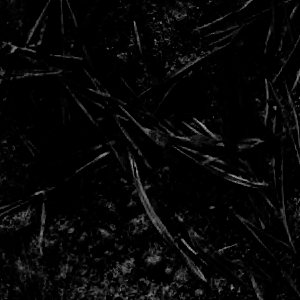} &
        \includegraphics[width=0.15\linewidth]{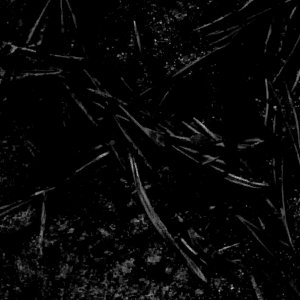} &
        \includegraphics[width=0.15\linewidth]{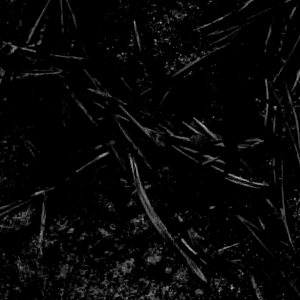} \\
    \end{tabular}
    \caption{Comparison of different methods for the \textit{forest\_sand\_01} material, retrieved from Poly Haven~\cite{polyhaven}.
    This material shows relatively lower PSNR and is one of the most difficult cases containing high-frequency details and different spatial patterns on textures.}
    \label{fig:forest}
\end{figure*}

Fig.~\ref{fig:roof} and Fig.~\ref{fig:forest} illustrate qualitative comparisons for all textures in the \textit{roof\_09} and \textit{forest\_sand\_01} materials, respectively.
For both materials, the naive approach produces block artifacts, while NTBC significantly reduces these artifacts by yielding a bit blurry images.
Conservative NTBC achieves good quality for the \textit{roof\_09} material, which shows about $10\%$ quality loss with about $67\%$ size reduction.
This material has a significant correlation across all textures, which makes it easier for our method to learn intrinsic features inherent in the material.
On the other hand, the \textit{forest\_sand\_01} material is one of the most challenging cases where all the textures have very high-frequency contents, and ARM, displacement, and specular textures seem to show different spatial features.
It results in similar results of NTBC to the naive approach even with the conservative approach, especially for a specular texture.

\section{Texture Dataset Details}
\begin{figure*}
    \centering
    \small
    \setlength{\tabcolsep}{0\linewidth}
    \renewcommand{\arraystretch}{0.6}
    \begin{tabular}{c@{\hspace{0.004\linewidth}} c@{\hspace{0.004\linewidth}} c@{\hspace{0.004\linewidth}} c@{\hspace{0.004\linewidth}} c@{\hspace{0.004\linewidth}} c@{\hspace{0.004\linewidth}} c}

        & Bricks090 & Carpet015 & MetalPlates013 & PavingStones070 & Rails001 & Wood063 \\

        \raisebox{0.08\linewidth}{\rotatebox[origin = c]{90}{Diffuse}} &
        \includegraphics[width=0.16\linewidth]{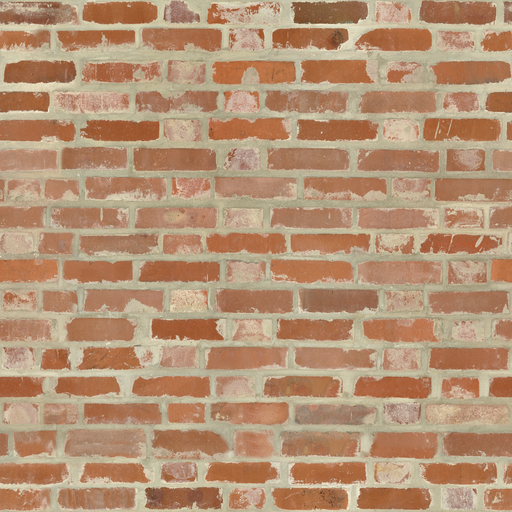} &
        \includegraphics[width=0.16\linewidth]{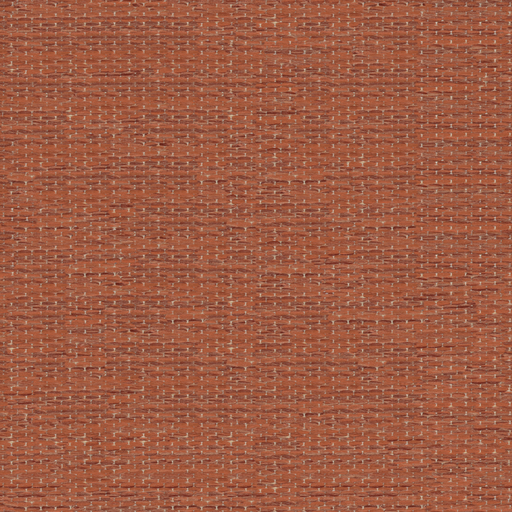} &
        \includegraphics[width=0.16\linewidth]{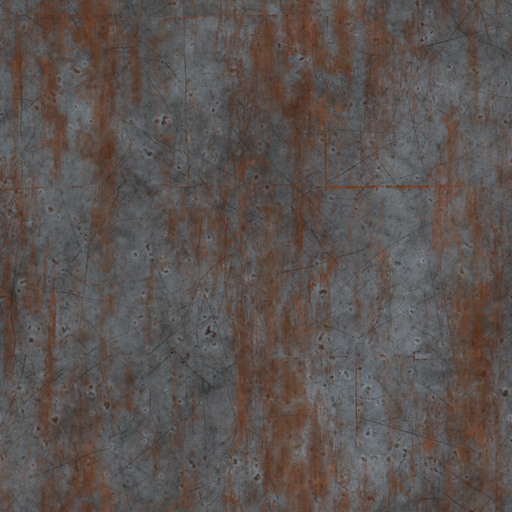} &
        \includegraphics[width=0.16\linewidth]{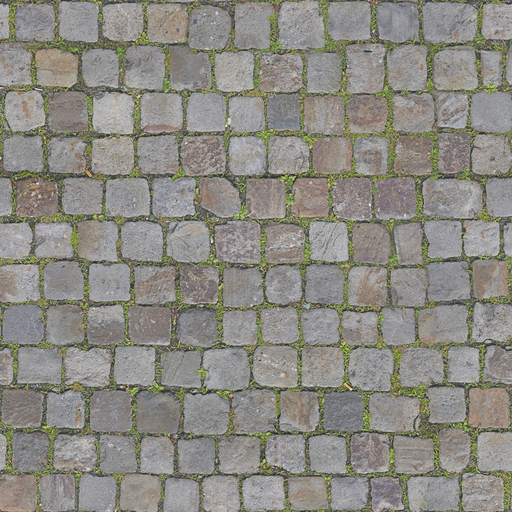} &
        \includegraphics[width=0.16\linewidth]{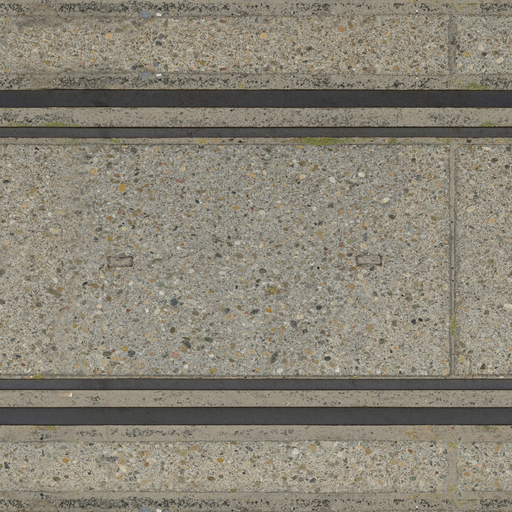} &
        \includegraphics[width=0.16\linewidth]{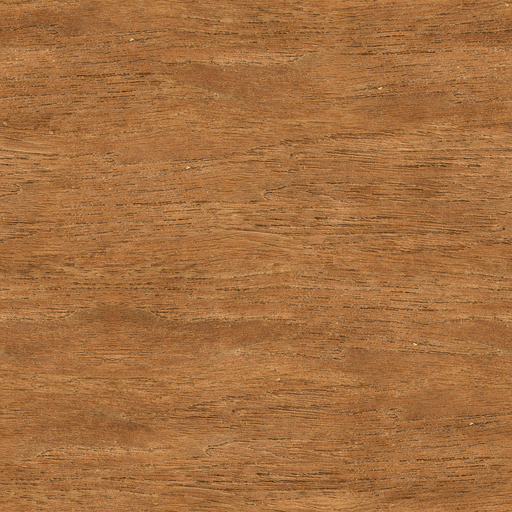} \\


        \raisebox{0.08\linewidth}{\rotatebox[origin = c]{90}{Normal}} &
        \includegraphics[width=0.16\linewidth]{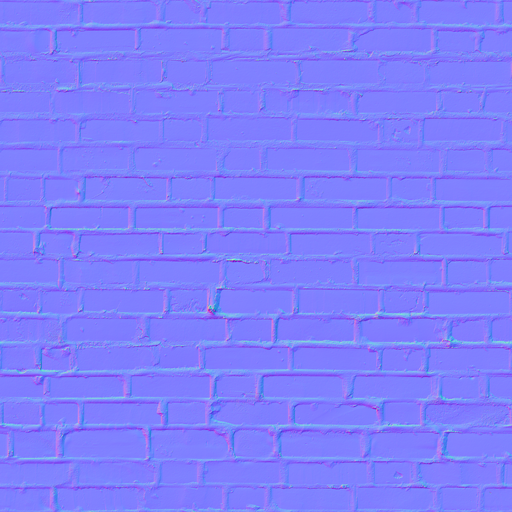} &
        \includegraphics[width=0.16\linewidth]{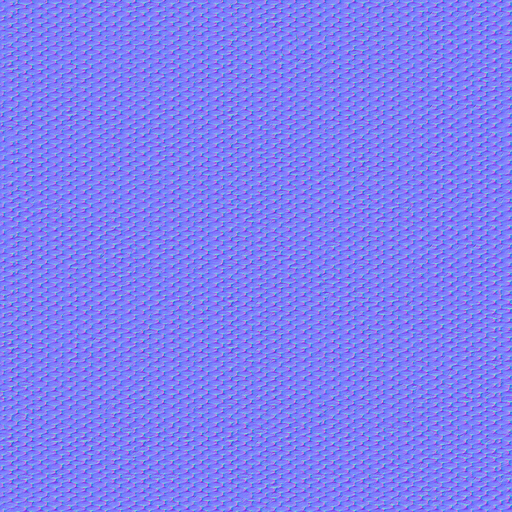} &
        \includegraphics[width=0.16\linewidth]{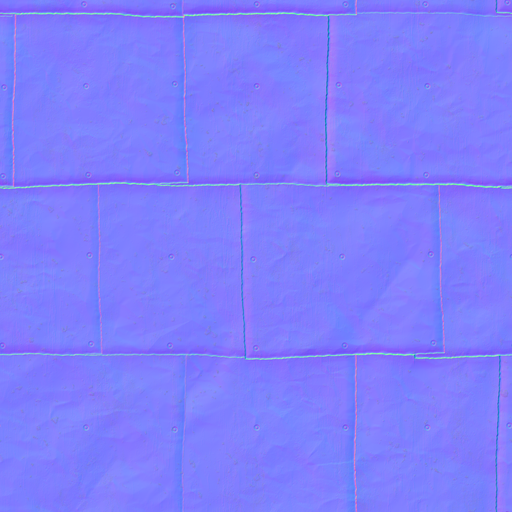} &
        \includegraphics[width=0.16\linewidth]{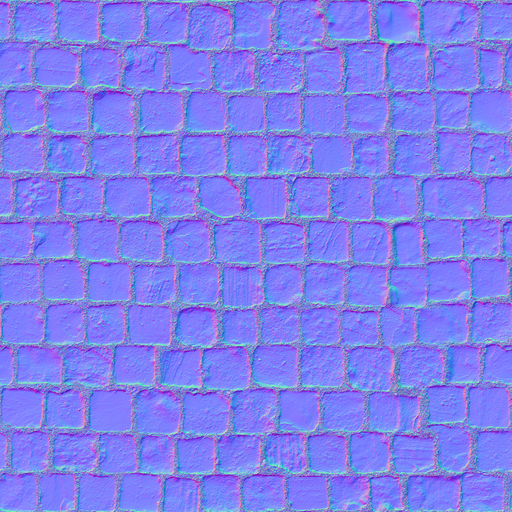} &
        \includegraphics[width=0.16\linewidth]{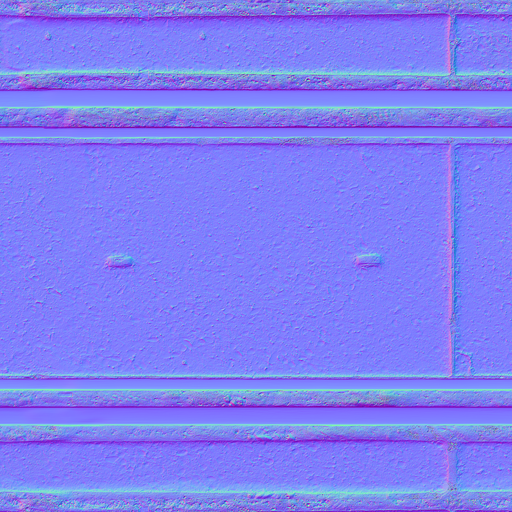} &
        \includegraphics[width=0.16\linewidth]{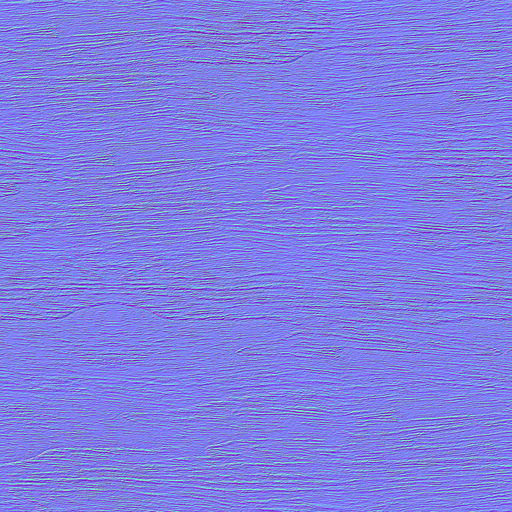} \\


        \raisebox{0.08\linewidth}{\rotatebox[origin = c]{90}{Ambient Occlusion}} &
        \includegraphics[width=0.16\linewidth]{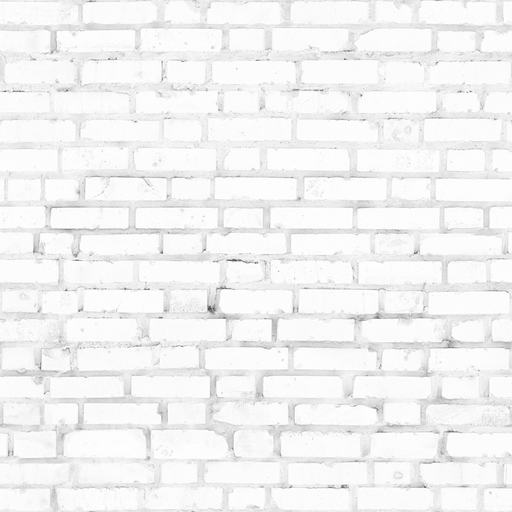} &
        \includegraphics[width=0.16\linewidth]{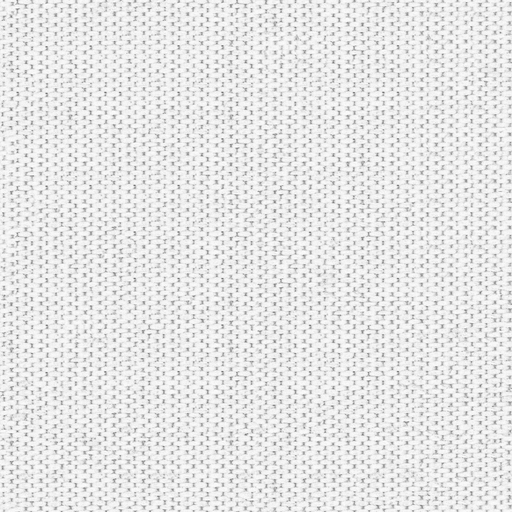} &
        \includegraphics[width=0.16\linewidth]{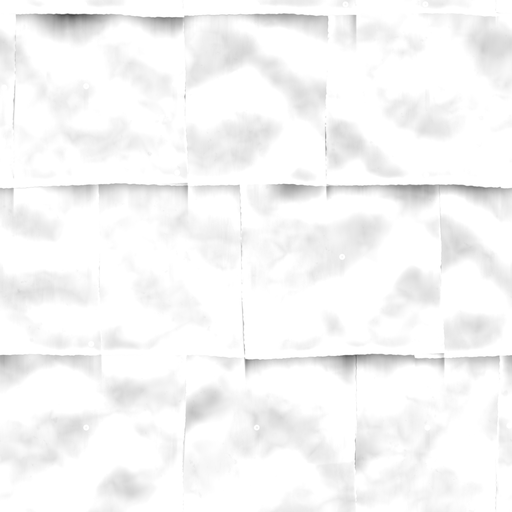} &
        \includegraphics[width=0.16\linewidth]{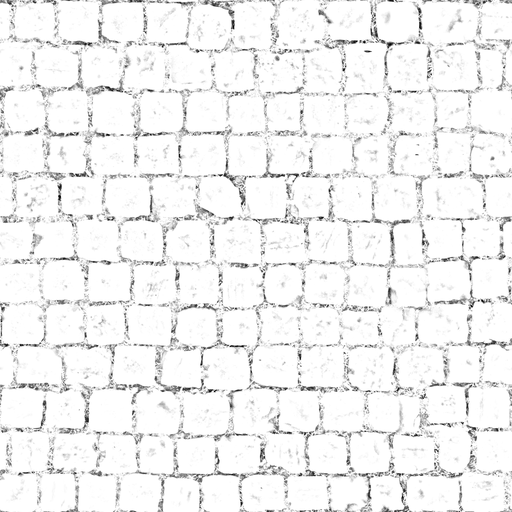} &
        \includegraphics[width=0.16\linewidth]{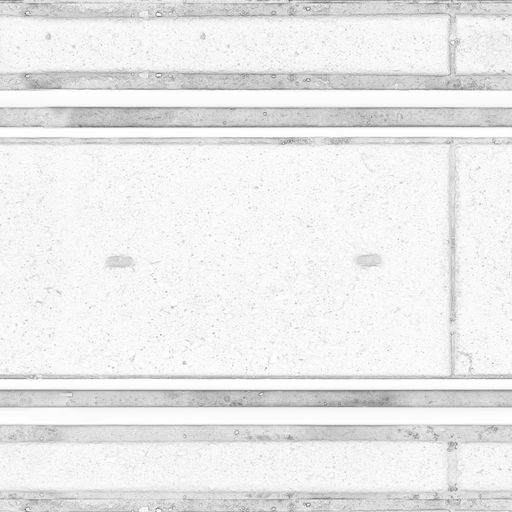} &
        \includegraphics[width=0.16\linewidth]{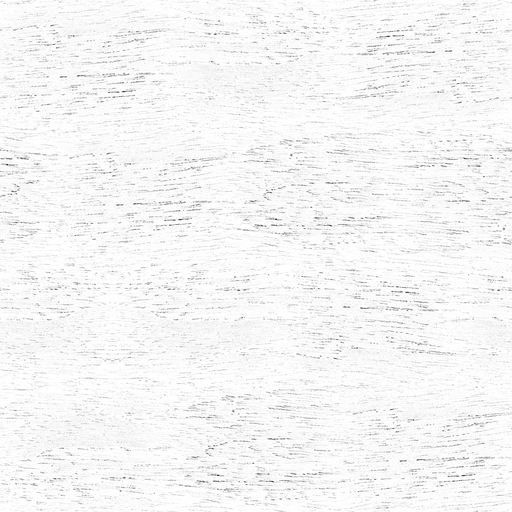} \\


        \raisebox{0.08\linewidth}{\rotatebox[origin = c]{90}{Displacement}} &
        \includegraphics[width=0.16\linewidth]{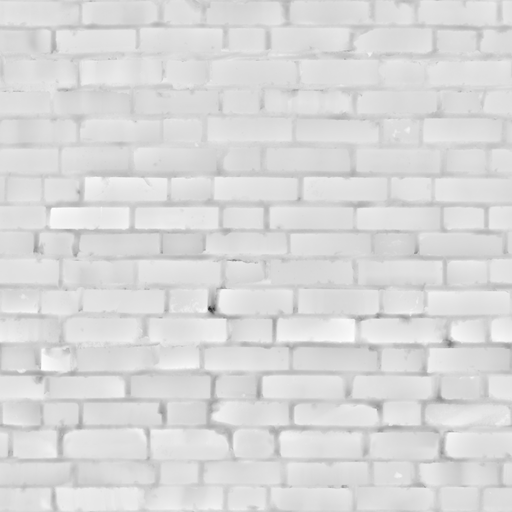} &
        \includegraphics[width=0.16\linewidth]{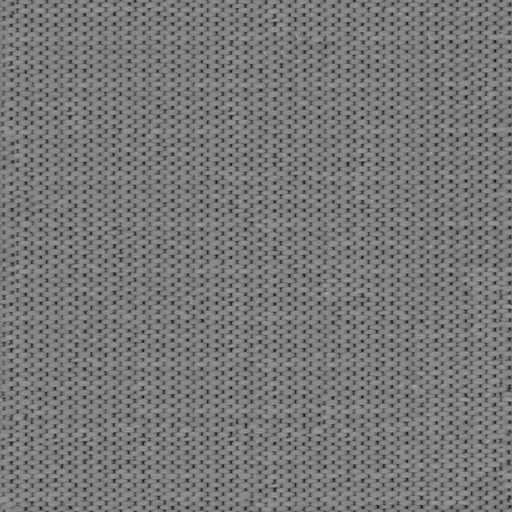} &
        \includegraphics[width=0.16\linewidth]{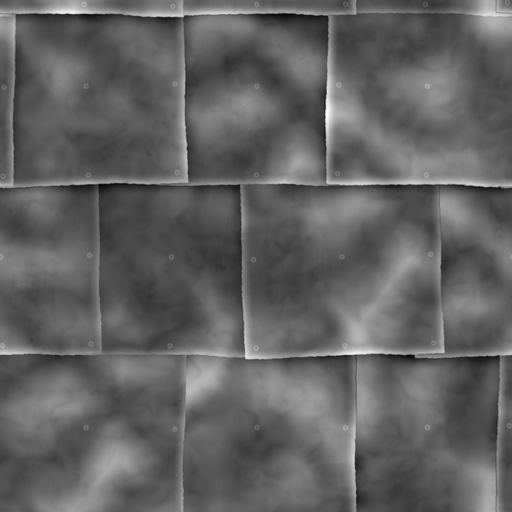} &
        \includegraphics[width=0.16\linewidth]{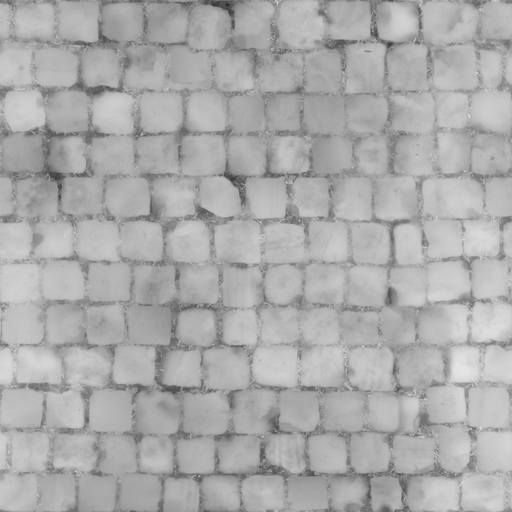} &
        \includegraphics[width=0.16\linewidth]{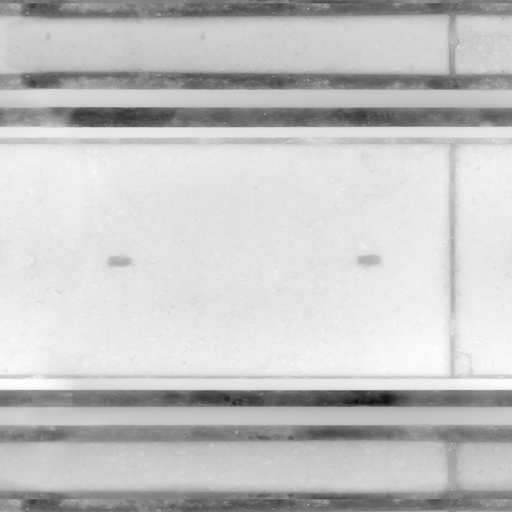} &
        \includegraphics[width=0.16\linewidth]{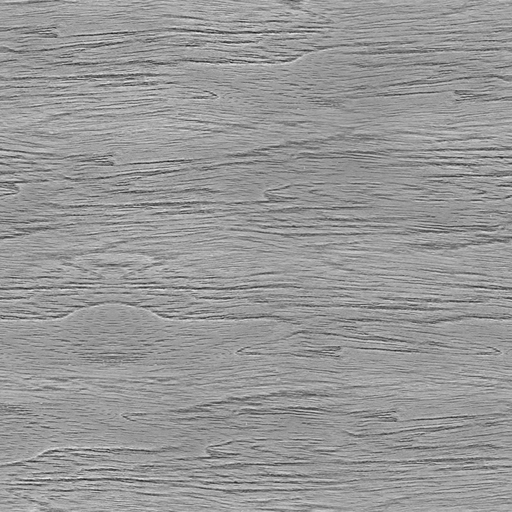} \\


        \raisebox{0.08\linewidth}{\rotatebox[origin = c]{90}{Roughness}} &
        \includegraphics[width=0.16\linewidth]{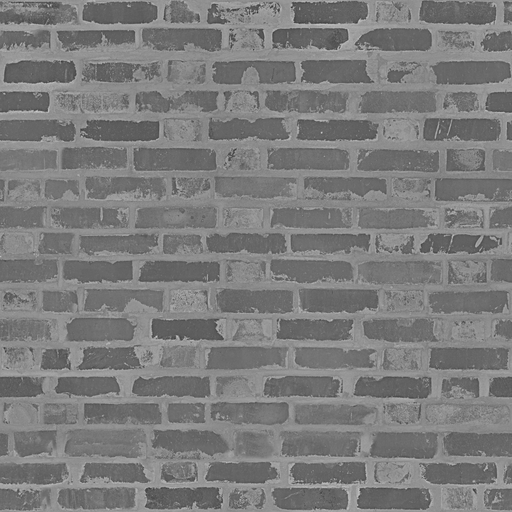} &
        \includegraphics[width=0.16\linewidth]{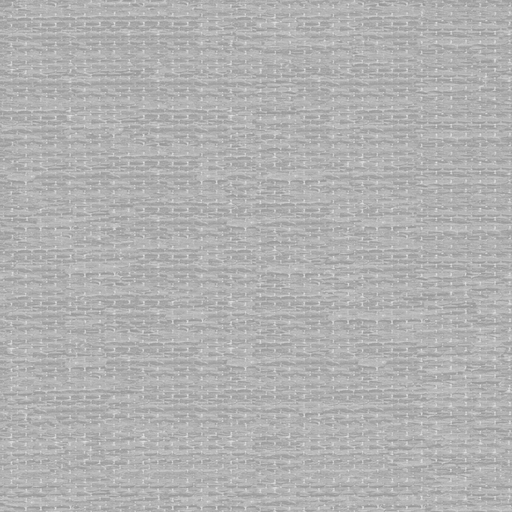} &
        \includegraphics[width=0.16\linewidth]{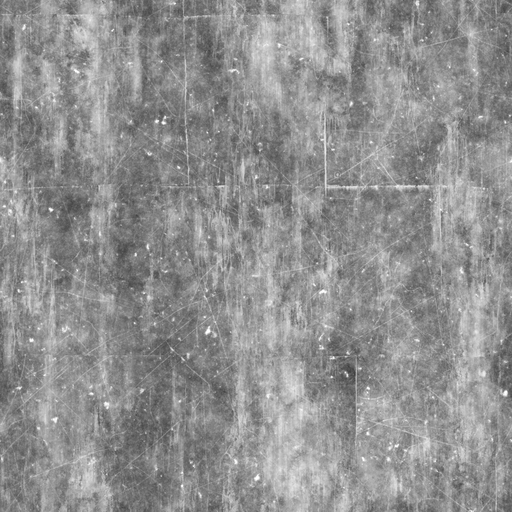} &
        \includegraphics[width=0.16\linewidth]{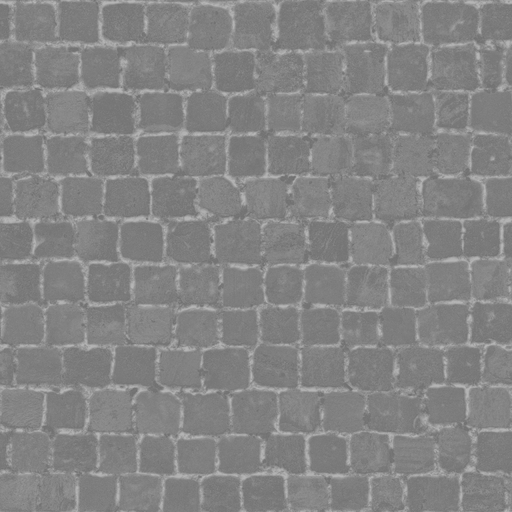} &
        \includegraphics[width=0.16\linewidth]{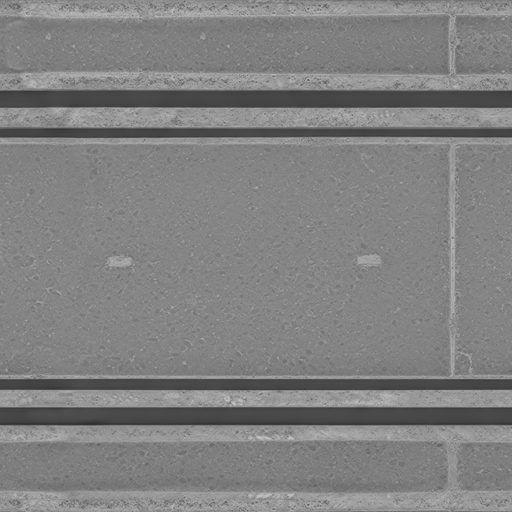} &
        \includegraphics[width=0.16\linewidth]{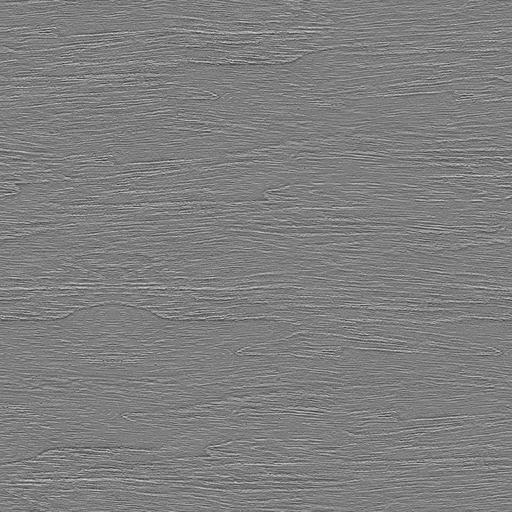} \\


        \raisebox{0.08\linewidth}{\rotatebox[origin = c]{90}{Metalness}} &
        & &
        \includegraphics[width=0.16\linewidth]{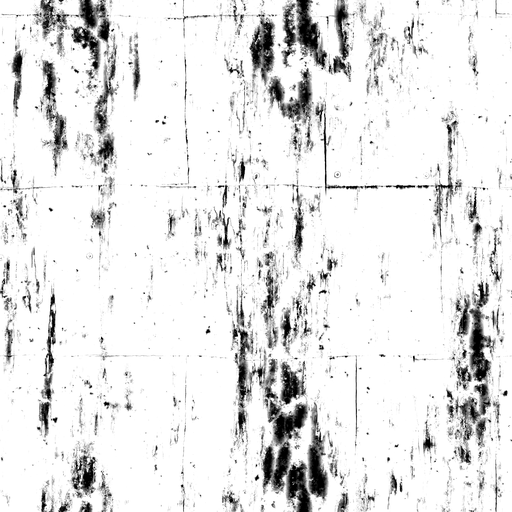} &
        &
        \includegraphics[width=0.16\linewidth]{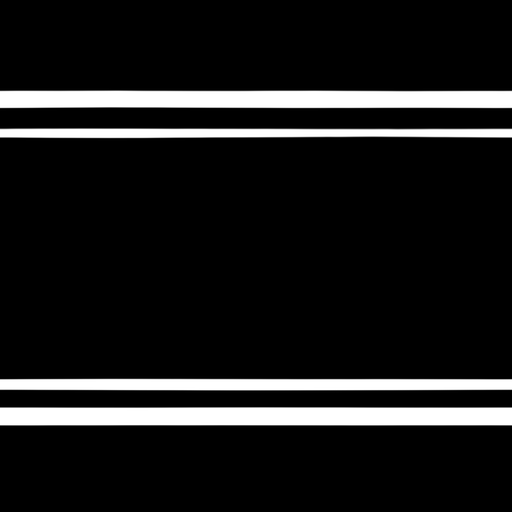} & \\
    \end{tabular}
    \caption{Materials retrieved from ambientCG~\cite{ambientCG}.}
    \label{fig:ambientCG}
\end{figure*}

\begin{figure*}
    \centering
    \small
    \setlength{\tabcolsep}{0\linewidth}
    \renewcommand{\arraystretch}{0.6}
    \begin{tabular}{c@{\hspace{0.004\linewidth}} c@{\hspace{0.004\linewidth}} c@{\hspace{0.004\linewidth}} c@{\hspace{0.004\linewidth}} c}

        & aerial\_rocks\_02 & forest\_sand\_01 & red\_dirt\_mud\_01 & roof\_09 \\


        \raisebox{0.08\linewidth}{\rotatebox[origin = c]{90}{Diffuse}} &
        \includegraphics[width=0.16\linewidth]{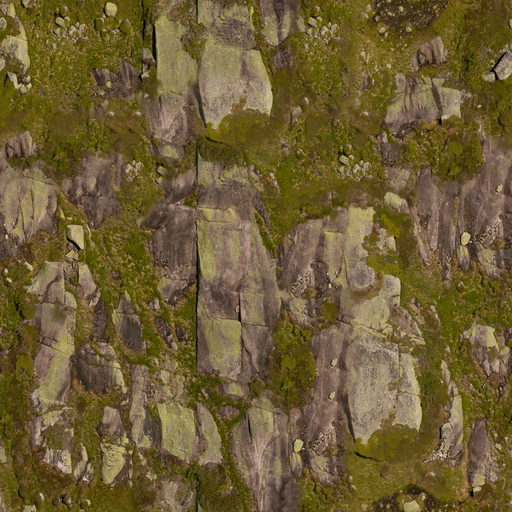} &
        \includegraphics[width=0.16\linewidth]{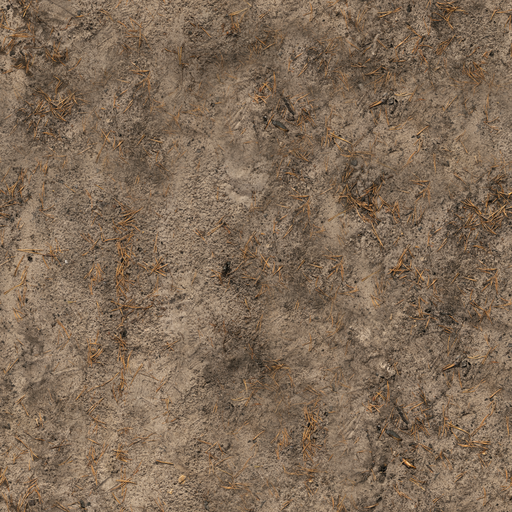} &
        \includegraphics[width=0.16\linewidth]{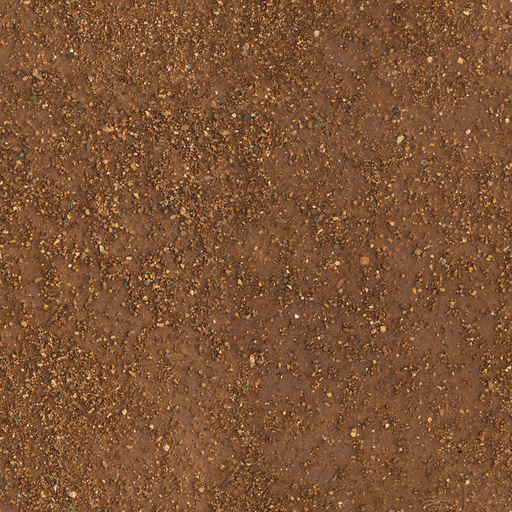} &
        \includegraphics[width=0.16\linewidth]{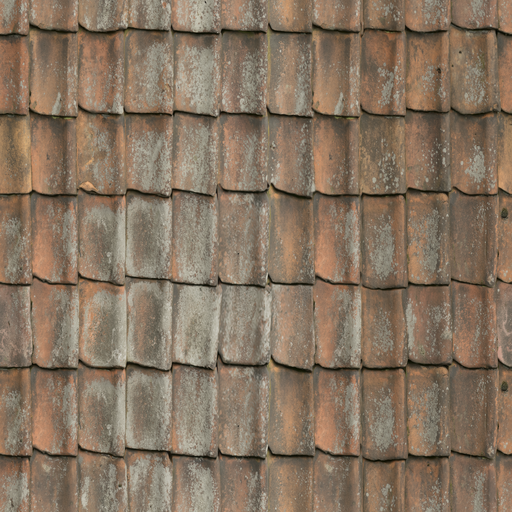} \\


        \raisebox{0.08\linewidth}{\rotatebox[origin = c]{90}{Normal}} &
        \includegraphics[width=0.16\linewidth]{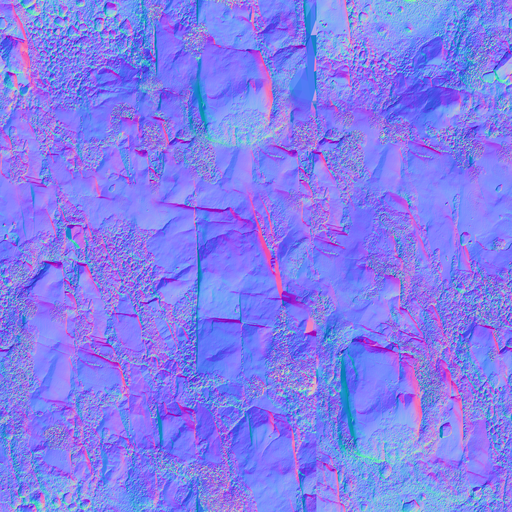} &
        \includegraphics[width=0.16\linewidth]{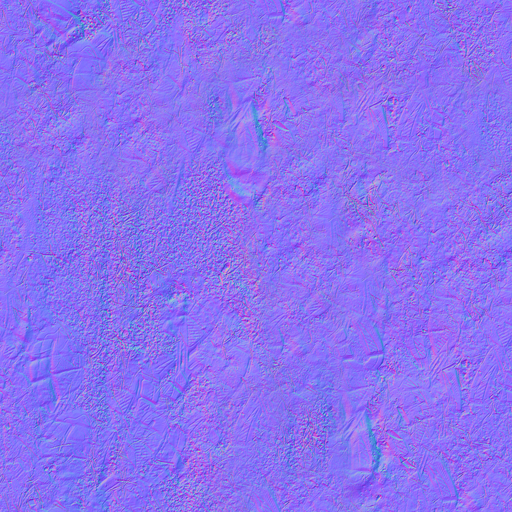} &
        \includegraphics[width=0.16\linewidth]{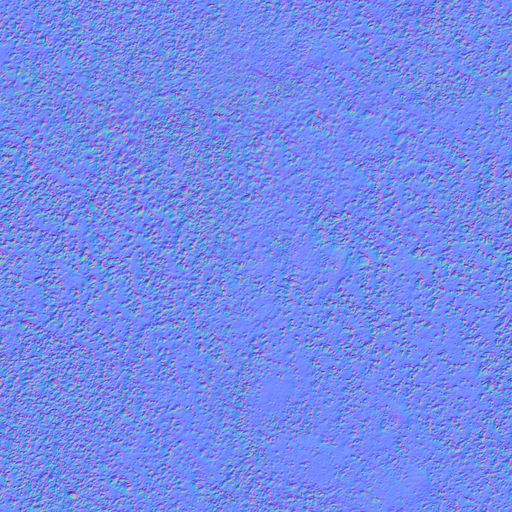} &
        \includegraphics[width=0.16\linewidth]{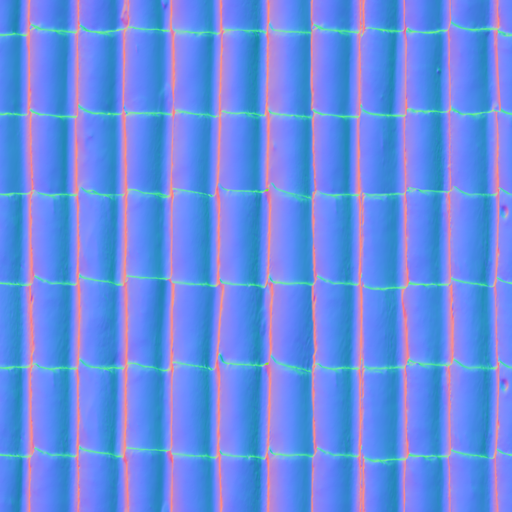} \\


        \raisebox{0.08\linewidth}{\rotatebox[origin = c]{90}{ARM}} &
        &
        \includegraphics[width=0.16\linewidth]{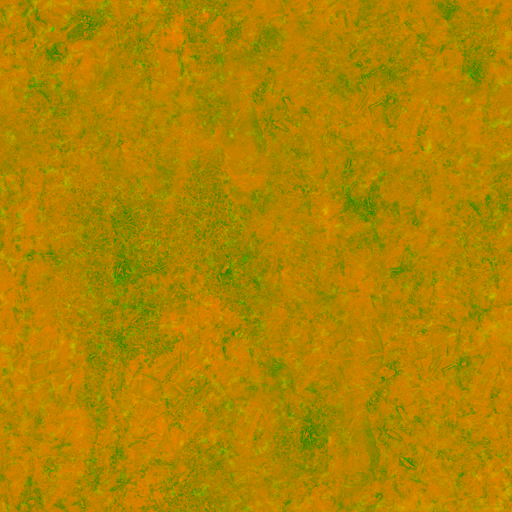} &
        \includegraphics[width=0.16\linewidth]{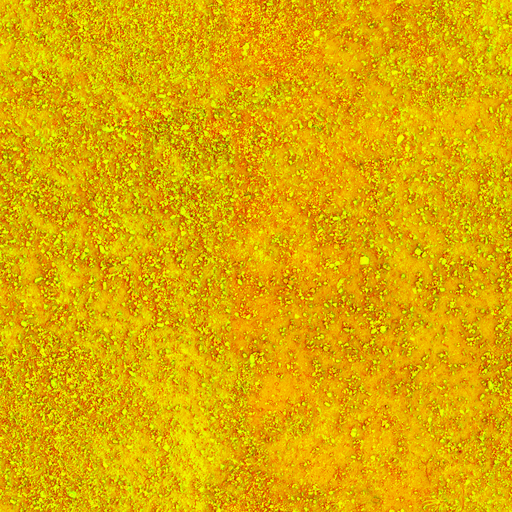} &
        \includegraphics[width=0.16\linewidth]{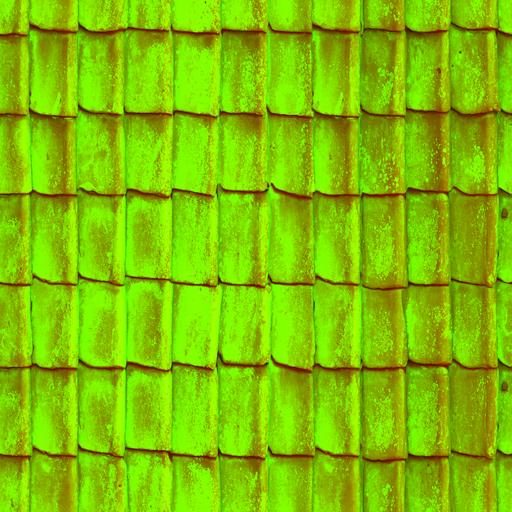} \\


        \raisebox{0.08\linewidth}{\rotatebox[origin = c]{90}{Displacement}} &
        \includegraphics[width=0.16\linewidth]{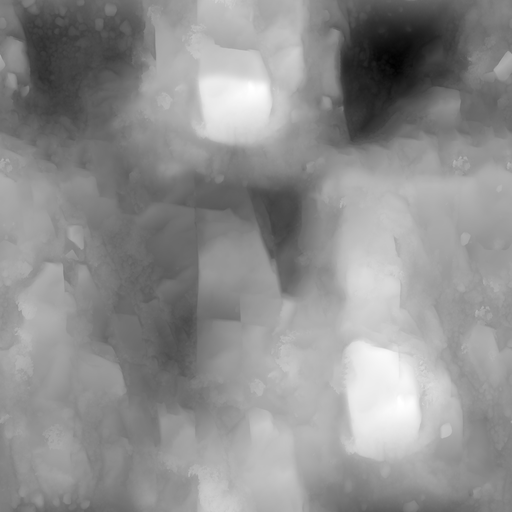} &
        \includegraphics[width=0.16\linewidth]{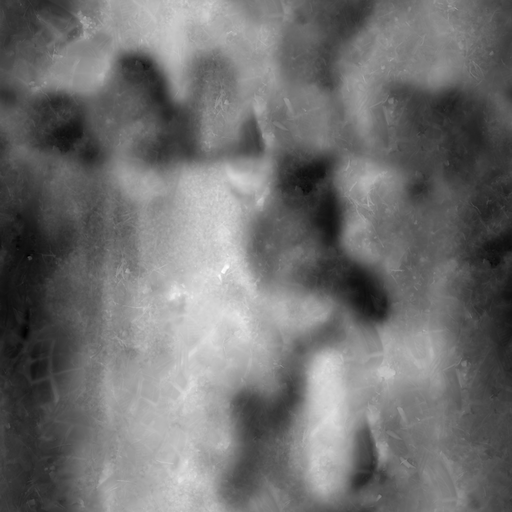} &
        \includegraphics[width=0.16\linewidth]{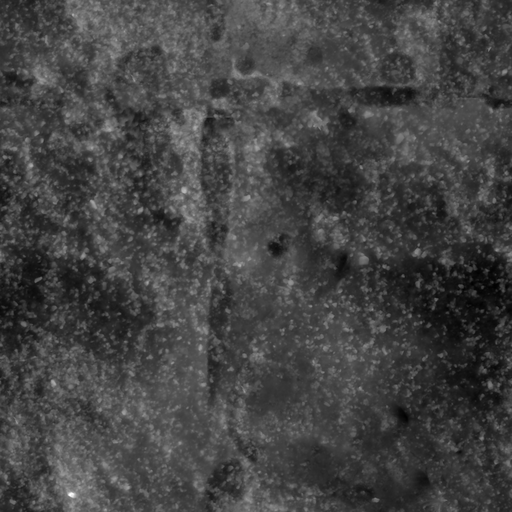} &
        \includegraphics[width=0.16\linewidth]{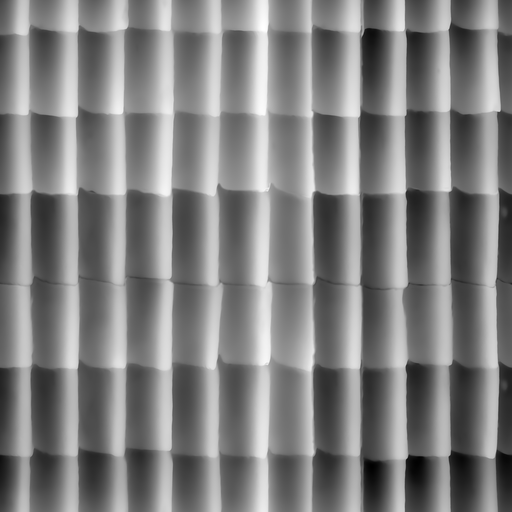} \\


        \raisebox{0.08\linewidth}{\rotatebox[origin = c]{90}{Specular}} &
        &
        \includegraphics[width=0.16\linewidth]{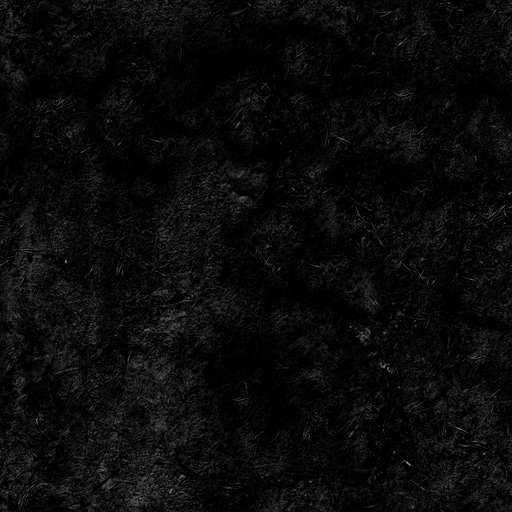} &
        \includegraphics[width=0.16\linewidth]{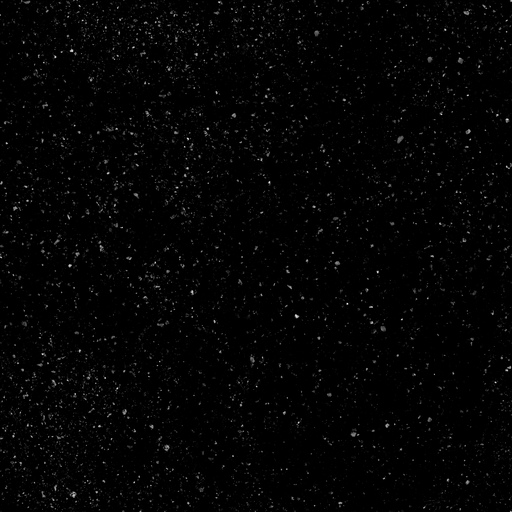} &
        \includegraphics[width=0.16\linewidth]{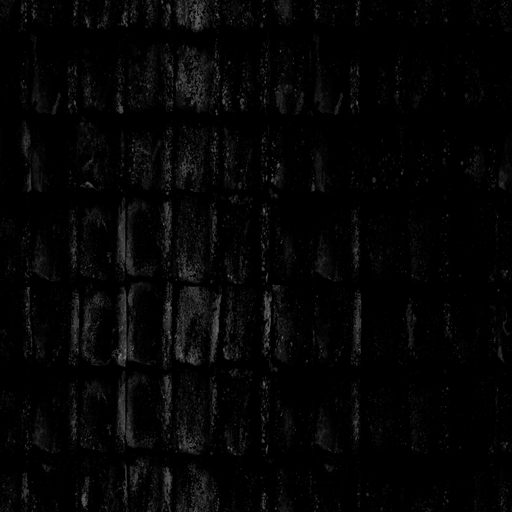} \\


        \raisebox{0.08\linewidth}{\rotatebox[origin = c]{90}{Ambient Occlusion}} &
        \includegraphics[width=0.16\linewidth]{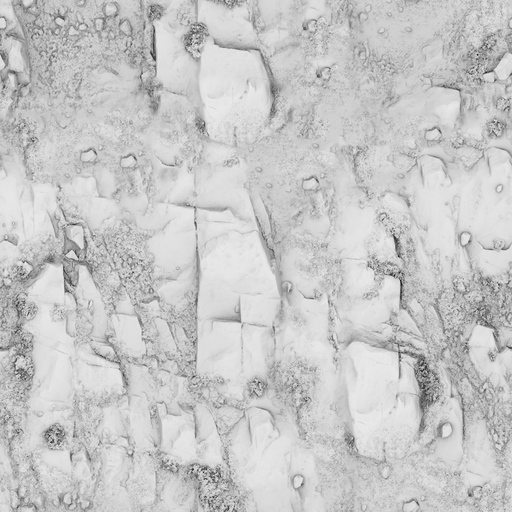} &
        & & \\


        \raisebox{0.08\linewidth}{\rotatebox[origin = c]{90}{Roughness}} &
        \includegraphics[width=0.16\linewidth]{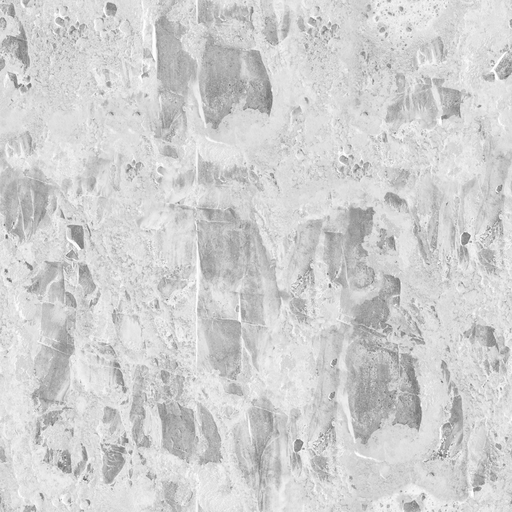} &
        & & \\
    \end{tabular}
    \caption{Materials retrieved from Poly Haven~\cite{polyhaven}.}
    \label{fig:PolyHaven}
\end{figure*}

Fig.~\ref{fig:ambientCG} and Fig.~\ref{fig:PolyHaven} show all textures of $10$ materials in our dataset, retrieved from ambientCG~\cite{ambientCG} and PolyHaven~\cite{polyhaven} websites.
All materials contain $5$ or $6$ textures with various numbers of RGB and single-channel textures.
All textures have a resolution of $4096 \times 4096$ pixels.
\textit{MetalPlates013} and \textit{Rails001} materials have $6$ textures, for which NTBC shows the best storage efficiency compared to the standard BC.
Our \textit{aggressive} and \textit{conservative} methods consume $13.37$ MB and $26.74$ MB, respectively, while the standard BC consumes $48$ MB, resulting in about $72\%$ and $45\%$ reduction in storage footprint.

\end{document}